\documentclass[12pt, preprint]{aastex}

\newcommand{\Msun}{\mbox{$M_\odot$}}
\newcommand{\Teff}{\mbox{$T_{\rm eff}$}}
\newcommand{\Logg}{\mbox{$\log g$}}
\newcommand{\FeH}{\mbox{[Fe/H]}}
\newcommand{\Mv}{\mbox{$M_{\rm V}$}}
\newcommand{\mrat}{\mbox{$M_1/M_2$}}

\shorttitle{Biased targets for Tycho 2}
\shortauthors{Ammons et al.}

\begin{document}

\title{N2K IV:  New Temperatures and Metallicities for 100,000+ FGK Dwarfs}

\author{S. Mark Ammons\altaffilmark{1,3}, Sarah E. Robinson\altaffilmark{1,4}, Jay Strader\altaffilmark{1,5}, Gregory Laughlin\altaffilmark{1,6}, Debra Fischer\altaffilmark{2,7}, Aaron Wolf\altaffilmark{1,8}}
\affil{Astronomy Department, University of California,
    Santa Cruz}

\altaffiltext{1}{present address:  Astronomy Department, University of California, Santa Cruz, 1156 High St., Santa Cruz, CA  95064}
\altaffiltext{2}{present address:  Physics and Astronomy, San Francisco State University, 1600 Holloway, San Francisco, CA 94132-4163}
\altaffiltext{3}{ammons@ucolick.org}
\altaffiltext{4}{ser@ucolick.org}
\altaffiltext{5}{strader@ucolick.org}
\altaffiltext{6}{laugh@ucolick.org}
\altaffiltext{7}{fischer@stars.sfsu.edu}
\altaffiltext{8}{kheldar@ucolick.org}

\begin{abstract}
We have created a framework to facilitate the construction of specialized target lists for radial velocity surveys that are strongly biased toward stars that (1) possess planets
and (2) are easiest to observe with current detection techniques.  We use a procedure that uniformly estimates fundamental stellar properties of Tycho 2 stars, with errors, using
spline functions of broadband photometry and proper motion found in Hipparcos/Tycho 2 and 2MASS.  We provide estimates of \Teff\ and distance for 2.4 million Tycho 2 stars that
lack trigonometric distances.  For stars that appear to be FGK dwarfs according to estimated \Teff\ and \Mv\, we also derive \FeH\ and identify unresolved binary systems with mass
ratios $1.25 <$ \mrat\ $< 3.0$.  Our spline function models are trained on the unique Valenti \& Fischer (2005) set, composed of 1000 dwarfs with precise stellar parameters
estimated from HIRES spectroscopy.  For FGK dwarfs with photometric error $\sigma_V < 0.05,$ or $V < 9$, our temperature model gives a one-sigma error of $\sigma_T = +58.7/-65.9$
K and our metallicity model gives a one-sigma error of $\sigma_{[Fe/H]} = +0.13/-0.14$ dex.  Addressing the need for lone stars in radial velocity surveys, the binarity model may be used to remove $70\%$ of doubles with $1.25 <$ \mrat\ $< 3.0$ from a magnitude-limited sample of dwarfs at a cost of cutting $20\%$ of the sample.  Our estimates of distance and spectral type enable us to isolate 354,822 Tycho 2 dwarfs, 321,996 of which are absent from Hipparcos, with giant and subgiant contamination at $2.6\%$ and $7.2\%,$ respectively.  100,000 of these stars are not in Hipparcos and have sufficiently low photometric errors to retain $0.13-0.3$ dex \FeH\ accuracy and $80-100$ K temperature accuracy (one-sigma).  2,500 of these FGK dwarfs are bright ($V < 9.0$) and metal-rich ($\FeH\ > 0.2$).  All estimates of stellar parameters published in this catalog are accompanied by accurate asymmetric errors that include propagated photometric error.  Our metallicity estimates have been used to identify targets for N2K (Fischer et al. 2005a), a large-scale radial velocity search for Hot Jupiters, which has verified the errors presented here.  The catalog of temperatures and metallicities that we derive can not only be used to construct candidate lists for targeted planet searches, but also to further large-scale studies of Galactic structure and chemical evolution and to provide potential reference stars for narrow angle astrometry programs such as SIM and large-aperture optical interferometry.  
\end{abstract}  


\section{INTRODUCTION}

The conventional reservoir of targets for radial velocity surveys has been the Hauck \& Mermilliod (1998) set of bright stars with $\it{uvby}$ narrow-band photometry available.  $\it{uvby}$ fluxes may be used to precisely estimate a dwarf's metallicity, [Fe/H], which is a good predictor of the presence of a planet (Valenti \& Fischer 2005).  As the Hauck \& Mermilliod reservoir has largely been screened for planets, it has become necessary to choose targets from larger surveys of dimmer stars.  Unfortunately, these expansive sets have lower-quality data from which to extract metallicity and few of these dim stars have distances available to sort out giants.  Nevertheless, even approximate estimates of metallicity and other fundamental stellar parameters can be used to assemble biased target lists for radial velocity surveys.  These lists will ultimately result in more positive detections and less time wasted on large-aperture telescopes.

Several of these large stellar catalogs are 2MASS
(Cutri et al. 2003), Tycho/Hipparcos (Perryman et al.
1997, Hog et al. 1998, 2000), and the SDSS DR1 (Strauss et al. 2002, Abazajian et al. 2003), which contain very good flux measurements for millions of stars.  While
fundamental stellar properties such as metallicity and effective temperature are best determined from high resolution spectroscopy,
broadband photometric estimates of these parameters are useful for
applications that profit more from large samples than precise information
for individual stars.  

Historically, photometric polynomial fits have been developed to facilitate
searches for certain classes of objects, like halo giants or subdwarfs.
\Teff\ models are common in the literature, as the relationship
between temperature and color is straightforward (see, e.g., Blackwell and Shallis 1977, Alonso et al.
1996, Montegriffo et
al. 1998, Alonso et al. 1999, Richichi et al. 1999, Houdashelt et al. 2000).  UV excess has long been used as a
proxy for metallicity (see, e.g., Sandage \& Walker 1955, Carney 1979, Cameron 1985, Karaali et al. 2003).  Much recent work has been done in
conjunction with the Sloan Digital Sky Survey (Lenz et al. 1998, Helmi et
al. 2003).  Of chief interest are the many polynomial fits that have been made for [Fe/H], using both broadband and narrowband fluxes (see Twarog 1980, Schuster and Nissan 1989a, Rocha-Pinto and Maciel 1996, Favata et al. 1997, Flynn and Morell 1997, Kotoneva et al. 2002, Martell and Laughlin 2002, Twarog et al. 2002).

Although many of these studies used stellar models to construct or constrain polynomial terms, we use an entirely empirical approach.  We use a training set of stars with both high
resolution spectra and broadband photometry to fit polynomials to the
broadband colors.  This set is from Valenti \& Fischer
(2005), which contains over 1000 F, G, and K dwarfs with Keck/HIRES
spectra. We fit polynomials and spline functions with a flexible $\chi^2$-minimization
procedure to $BV$ photometry from Hipparcos, Tycho 2, and the UBV
Photoelectic catalogs (Mermilliod
1987, 1994, Perryman et al. 1997, Hog et al. 2000), $JHK$ photometry from 2MASS (Cutri et al. 2003), and proper
motion (when available) from Hipparcos and Tycho 2.  The size and quality
of our training set and broadband database distinguish the present work
from previous studies.  

We estimate \Teff\ and distance for 2.4 million stars in Tycho 2. A subset of
$354,822$ FGK dwarfs also have estimates of \FeH\ and the probability of multiplicity.  These data have been concurrently published in electronic form and will be publicly available at the Astronomical Journal Supplement.  A primary purpose of
this work is to facilitate the selection of metal-rich FGK dwarfs for N2K, a radial
velocity survey of 2000 stars for hot Jupiters (Fischer et al. 2005a).  We also wish to isolate and remove stars for which it is difficult to obtain good radial velocities.  Toward this end, we demonstrate that one can construct an input list that is optimally free of subgiants and giants, early-type stars (O, B, A, and early F's), late-type stars (late K and M's), certain types of spectroscopic binaries, and metal-poor stars.  Although we do not directly publish a target list for radial velocity surveys, the published model estimates may be used to construct biased target lists through a ``figure of merit'' function of temperature, metallicity, distance, and probability of multiplicity.  In particular, the N2K project has used the \FeH\ estimates for Hipparcos stars to sort previous target lists.  Using follow-up spectroscopy (Robinson et al. 2005), future papers from the N2K consortium will confirm the success of our metallicity
estimates for a subset of these stars.

The layout of the paper is as follows.  \S 2 describes our least-squares
fitting program, error sources and estimation, and verification.  \S 3
contains the polynomial fits to the training set and errors.  \S 4 describes the application of
the polynomials to Tycho 2, deriving a pool of FGK dwarfs with \FeH\
estimates.  \S 5 discusses sources of error in the models and
improvements taken to address these errors.  \S 6 has our discussion and
conclusions. 

\section{METHOD}

\subsection{Multivariable Least-squares Fitting Approach}

Our models consist of $\chi^2$-optimized fitting polynomials on specialized training sets as described above.  These stars have measurements of a single ``training parameter,'' $B$ (i.e., \FeH, temperature, etc.), as well as a set of $N$ ``fitting parameters'' ${A_j}$ (i.e., photometry and proper motion).  Ideally, both the training parameter and fitting parameters are well-measured and their distributions are even across the range of each variable in the training set.  Our fitting routine first constructs the terms of a polynomial $f(A_1,A_2,A_3,\dots,A_N)$ from the fitting parameters $A_1,A_2,A_3,\dots,A_N.$  The coefficients of the terms of $f$ form a parameter space that may be minimized with respect to the least squares error $(f(A_1,A_2,A_3,\dots,A_N)-B)^2.$  The $\chi^2$ statistic takes the form
\begin{equation}
\chi^2 = \sum_{j}^{P} \left( \frac{B_j - f(A_{j1}, A_{j2}, \dots, A_{jN})}{\sigma_{B_j, A_j}} w_j \right)^2,
\end{equation}
where $P$ is the number of stars in the training set, $B_j$ is the training parameter for a particular star $j$, $A_{j1}, A_{j2}, \dots, A_{jN}$ is the set of fitting parameters for that star, $w_j$ is
a weight term, and $\sigma_{B_j,A_j}$ is some Gaussian measure of the error in both the training parameter and the set of fitting parameters.  Our fitting routine uses the Levenberg-Marquardt minimization scheme (e.g., Press et al. 1992) to optimize the coefficients of the terms in $f$ with respect to $\chi^2.$  The terms themselves are constructed to be the most
general up to a specified polynomial order $Q$.  The routine uses every existing permutation of variable exponents in the set of terms it generates, excluding those with a sum of exponents exceeding $Q.$

The exact assignment of the Gaussian error $\sigma_{B_j, A_j}$ in $\chi^2$ is nontrivial, and its form is essential for properly reducing the weight of stars with uncertain measurements.  We use a general form of $\sigma_{x_j,y_j}^2$ for multivariable, nonlinear fits:
\begin{equation}
\sigma_{B_j, A_j}^2 = \sigma_{B_j}^2 + \left( \sum_{j}^{P} \frac{\partial f^*}{\partial A_j}\sigma_{A_j} \right)^2,
\end{equation}
where the function $f^*$ is the polynomial found by a prior iteration of the routine.  In this prior iteration, $\chi^2$ is constructed according to equation (1) with an error $\sigma_{B_j, A_j}$ that has $f = 0.$

It must be noted that the training set is composed of bright stars, so the resulting models we present are optimized for stars with the ``best'' data.  This can be changed by using ``photometry-optimized'' models that employ a modified $\chi^2$ statistic that includes Gaussian photometry error in the fitting parameters.

\subsection{Three Way Data Split}
Our empirical broadband models are functions of $B_T$, $V_T$, $J$, $H$, and $Ks$ fluxes and proper motion.  These polynomials are very blunt tools, so they are best applied in a statistical sense to large numbers of stars.  They possess relatively large errors compared to spectroscopic or narrowband photometry models, and a thorough understanding of these errors is necessary.  We use well-documented techniques from machine learning and statistical data modeling to prevent our error estimates from being biased.  Cross-validation methods employ a splitting of the data set into one or more groups, some of which are used to generate model coefficients and others of which are used to estimate error; for a survey of these schemes see Weiss and Kulikowski (1991).  The most simple form of cross validation is a 50/50 split into a training set and a test set.  

In our application, we have an extra model validation step that necessitates an extra split in the data beyond 50/50.  We tune parameters in the fitting procedure as well as evaluate which types of model are best to use.  These tunings and choices must be performed by estimating the error with a separate data set, one other than the final test set used to derive the published error.  This prevents hidden correlations and biases from affecting the final error estimates.  This intermediate validation step is performed with a ``validation set.''  The three principal model tunings we make are as follows:
\begin{enumerate}
\item[(1)]Decide whether to find model coefficients by minimizing $\chi^2$ as shown in equation (1) or the sum of the absolute deviations, obtained by taking the square root of the quantity in the sum before performing the sum.  We have modified the Press et al. (1992) version of Levenberg-Marquardt minimization to allow this.  Minimizing the absolute deviation reduces the effect of outliers.  
\item[(2)]Choose the number of temperature bins.  The models are spline functions, with different polynomial fits for different temperature ranges (see \S 2.3).
\item[(3)]Choose the order of the polynomials.  For stability's sake, polynomial orders greater than 2 are not considered.  
\end{enumerate}

The method we have described is the three-way data split, in which the data set is divided into a ``training set,'' a ``validation set,'' and a ``test set.''  The steps we follow are:
\begin{enumerate}
\item[(1)]Compose many different models using permutations of the three model tunings enumerated above.  Train these models and generate model coefficients for each one by minimizing $\chi^2$ on the ``training set.''  
\item[(2)]Test each of the models on the validation set and compare the widths of the residuals.  We choose the final model permutation with the lowest one-sigma errors over a specified range of polynomial output.  
\item[(3)]Using the best model parameters and tunings, generate the final model coefficients by minimizing $\chi^2$ or the absolute deviation on the combination of the ``training set'' and the ``validation set.'' 
\item[(4)]Run the final model fit on the test-set stars and find the one-sigma asymmetric errors from the residuals.  These errors are later combined with propagated photometry error to give the final errors.
\end{enumerate}

\subsection{Model Construction}

Using Hipparcos stars as training sets, we have constructed a \Teff\ polynomial model, a distance model, a ``binarity'' model that makes a rough estimate of the probability of multiplicity, and a \FeH\ model for FGK type dwarfs.  The polynomials used in each case are functions of five colors:  $B_T-V_T$, $B_T-J$, $V_T-H$, $V_T-K$, and $J-K$.  The distance and \FeH\ models additionally input the total proper motion in mas/year.  The colors were chosen to use each of the Tycho $B_T$ and $V_T$ and 2MASS $J,$ $H,$ and $Ks$ magnitudes at least once and to minimize the number of color sums and differences necessary to arrive at any permutation, using the most sensitive colors to physical parameters.  Colors were used instead of apparent magnitudes to prevent distance-related observational biases from affecting the results.

The temperature models are simple polynomial functions of the fitting parameters above.  For distance, binarity, and metallicity, we use multi-dimensional unconnected splines.  The color variation for these stellar parameters is dependent on the temperature in complicated ways.  Simple high-order polynomials cannot capture these effects, but splines are capable of weighting colors differently as the temperature varies.  Also, splines are well-behaved beyond the edge of the $A_j$ space populated by the training set and are robust to outliers, unlike simple high order polynomials.  We create the spline by dividing the training set into many temperature bins and generating individual polynomials for each bin.  The temperature boundaries are chosen to keep equal numbers of stars in each bin.  Each polynomial in the spline is discontinuous with the polynomials in the neighboring temperature bins, but we perform a pseudo-connection \emph{a posteriori} as described in \S 5 with equation (5).  

As described in \S 2.2, the polynomial orders are selected in each case by minimizing the residual error in the model validation step.  The number of temperature bins is selected similarly, with separate polynomials generated for stars in each bin.  Polynomial coefficients can either be found by minimizing the absolute deviation or minimizing $\chi^2;$ this choice is also made during the model validation step.  Each of the training set stars are weighted with equation (3), where $f^*$ is found via a previous iteration that sets $\sigma^2_{B_j, A_j} = \sigma^2_{B_j}.$  The explicit weight terms $w_j$ in the sum are set to unity.  

\subsection{Sources of Model Error}

Most of the sources of modeling error that affect the quality of the polynomial are reflected in the error distributions obtained from the three-way data split, and are well-understood.  See \S 5 for source-by-source explanations.  Apart from these errors, the polynomial itself may be suboptimal due to the difficulties of finding absolute minima in the $\chi^2$ space.  Levenberg-Marquardt searches large variable spaces well, but the combination of noisy data and overly large ($\sim 50$ variables) spaces reduces its effectiveness.  Generally, the procedure stumbles upon one of many degenerate local minima.  The residuals and $\chi^2$ associated with each of these minima are comparable, and their estimates of the training parameter for the training set stars seem to agree to within the measurement error $<\sigma_{B_j, A_j}>.$  The polynomial coefficients in low variable spaces $(\sim 15)$ appear to be independent of the initial first guess as well.  Fortunately, even if the model polynomial is suboptimal for the given data set, the errors found by the three-way data split are accurate. 

One large source of error that escapes the three-way data split is a lack of similarity between the ``application set'' and the training set.  Here we refer to the ``application set'' as the group of stars upon which the model is applied, i.e., Tycho 2 for the present study.  Any polynomial is optimized for the regions of the $A_j$ space that are overpopulated by the training set.  The total $\chi^2$ for all stars is less affected by rare stars and the errors are consequently higher for the $A_j$ regions that they occupy.  Most importantly, the calculated errors in these regions are themselves very uncertain due to small-number statistics.  If the locations of overdense regions in the $A_j$ space of the application set are much different from those of the training set, the introduced errors are unknown.  Catalog selection effects are the dominant contribution to this error.  For example, the completeness limits of Hipparcos and Tycho 2 are $V \sim 7.5$ and $V \sim 11.0,$ respectively, and the histograms of apparent magnitudes within the catalogs have different peak locations.  The distribution of true distance varies accordingly, as do the distributions of spectral type and luminosity class.  Reddening plays a larger role in Tycho and may cause contamination issues from which Hipparcos is free.  These effects have not been quantified for this paper, although corrections can be made with prior knowledge about the selection effects.  Actual spectroscopic observation of the training parameter for random stars within the application set would reveal any of these systematic offsets.  

Fortunately, the training set stars we use may all be found in Hipparcos.  Hipparcos is ``embedded'' in Tycho 2, the application set, meaning that a single instrument was used to reduce the photometry and proper motion of both sets.  If this were not the case, additional systematic errors would be introduced that would be difficult to quantify or even identify.  It must be mentioned that these types of errors have historically been the roadblocks preventing widespread use of broadband \FeH\ models, as it is difficult to observe a uniform training set.

\section{RESULTS}

\subsection{Effective Temperature}

Surface temperature measurements from SME processing in the Valenti \& Fischer set (2005) have one-sigma errors of $44$ K for FGK dwarfs.  We have trained a temperature model on this data, which gives an accurate polynomial for dwarf sets without giant contamination.  However, such a model cannot be applied to an arbitrary magnitude-limited sample without unquantified giant contamination error, so we have created a separate ``coarse'' temperature model that trains on a set of both dwarfs and giants.  This larger set of 2433 stars includes dwarf and giant temperatures from the Cayrel de Strobel et al. (1997) compilation as well as 852 Valenti \& Fischer dwarfs.  The temperature measurement error is given a value of $\pm 100$ K for the Cayrel de Strobel stars, which is the level of scatter we find for multiple measurements of single stars in this set.  The errors and parameters for both the coarse and fine models are given in Table 1.  The test scatter plots and error histograms are shown in figure 1.  

\begin{figure}[htb]
\epsscale{1.0}
\centerline{\plottwo{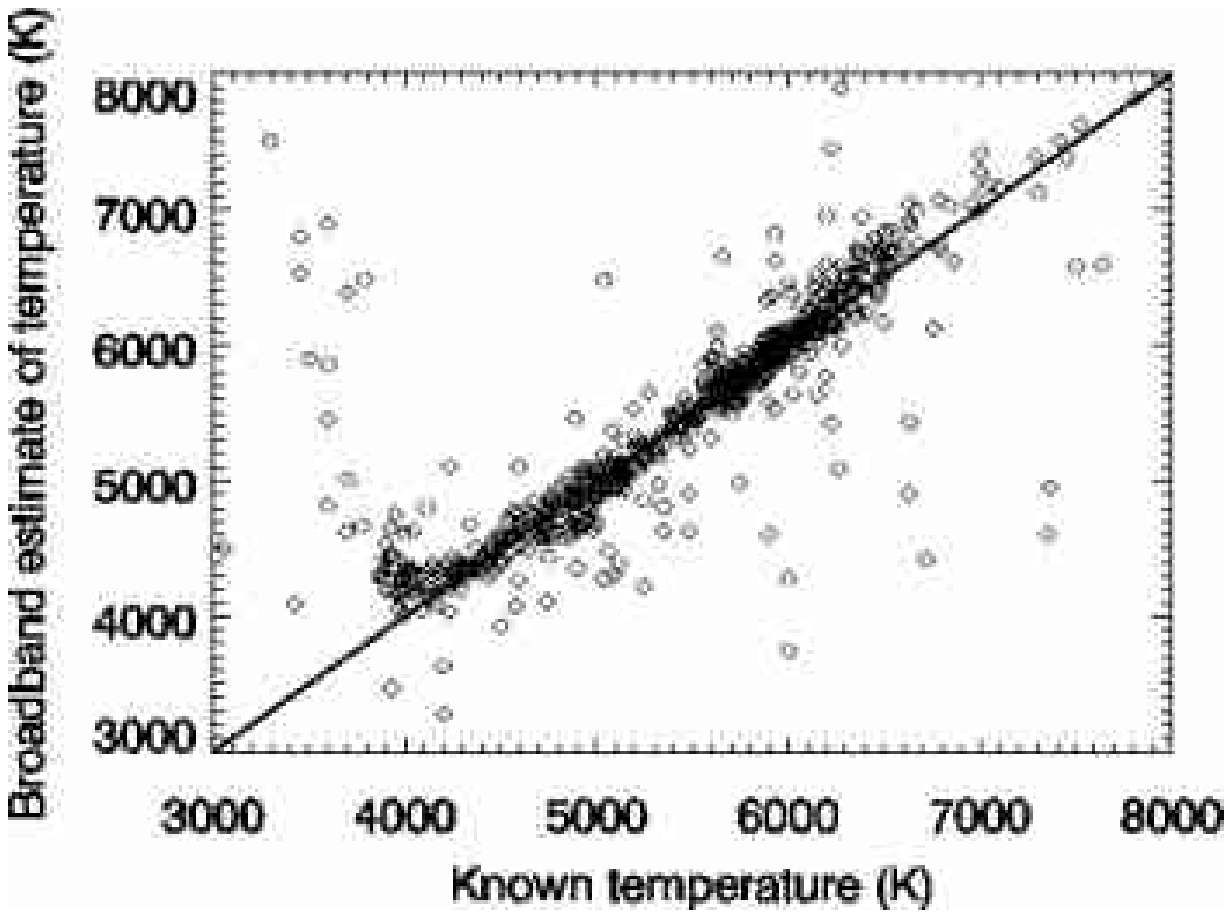}{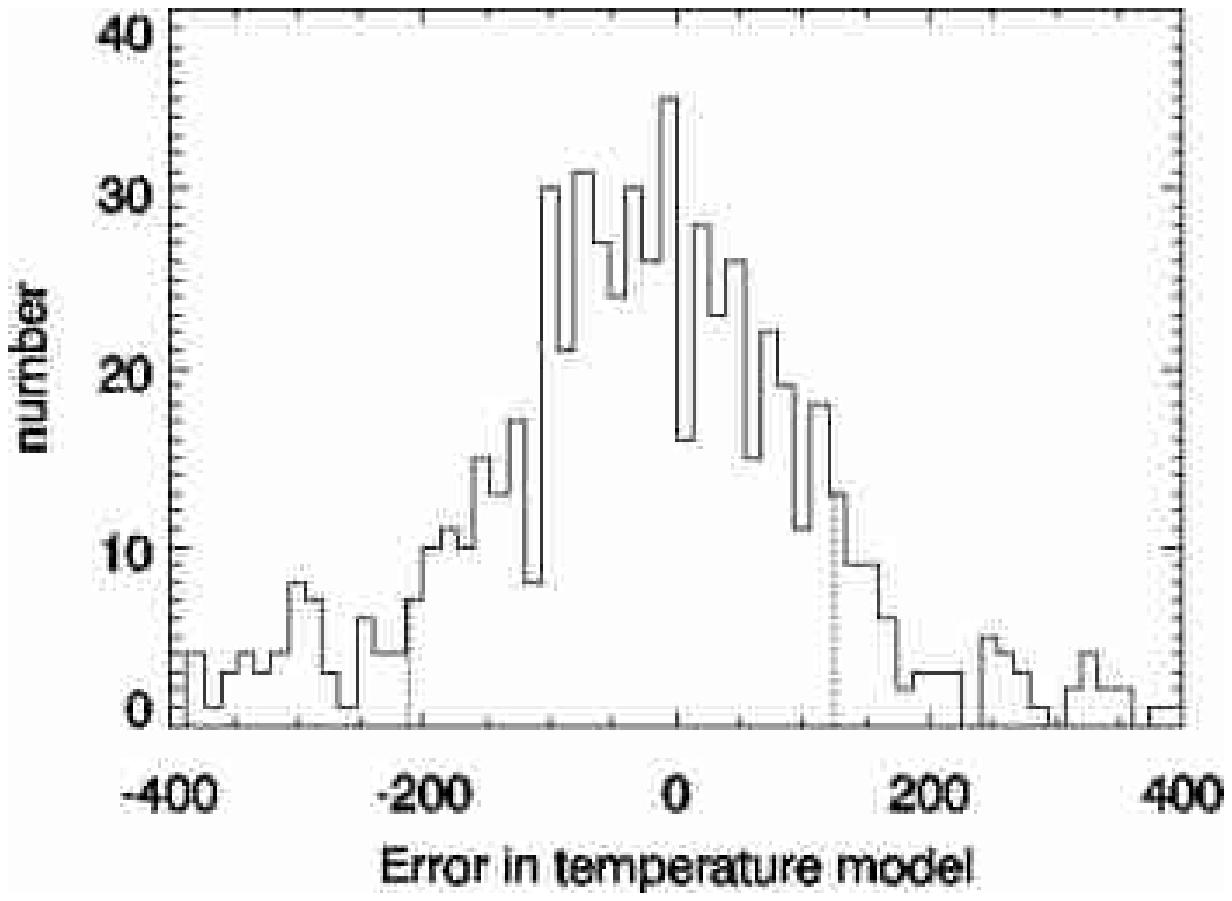}}
\centerline{\plottwo{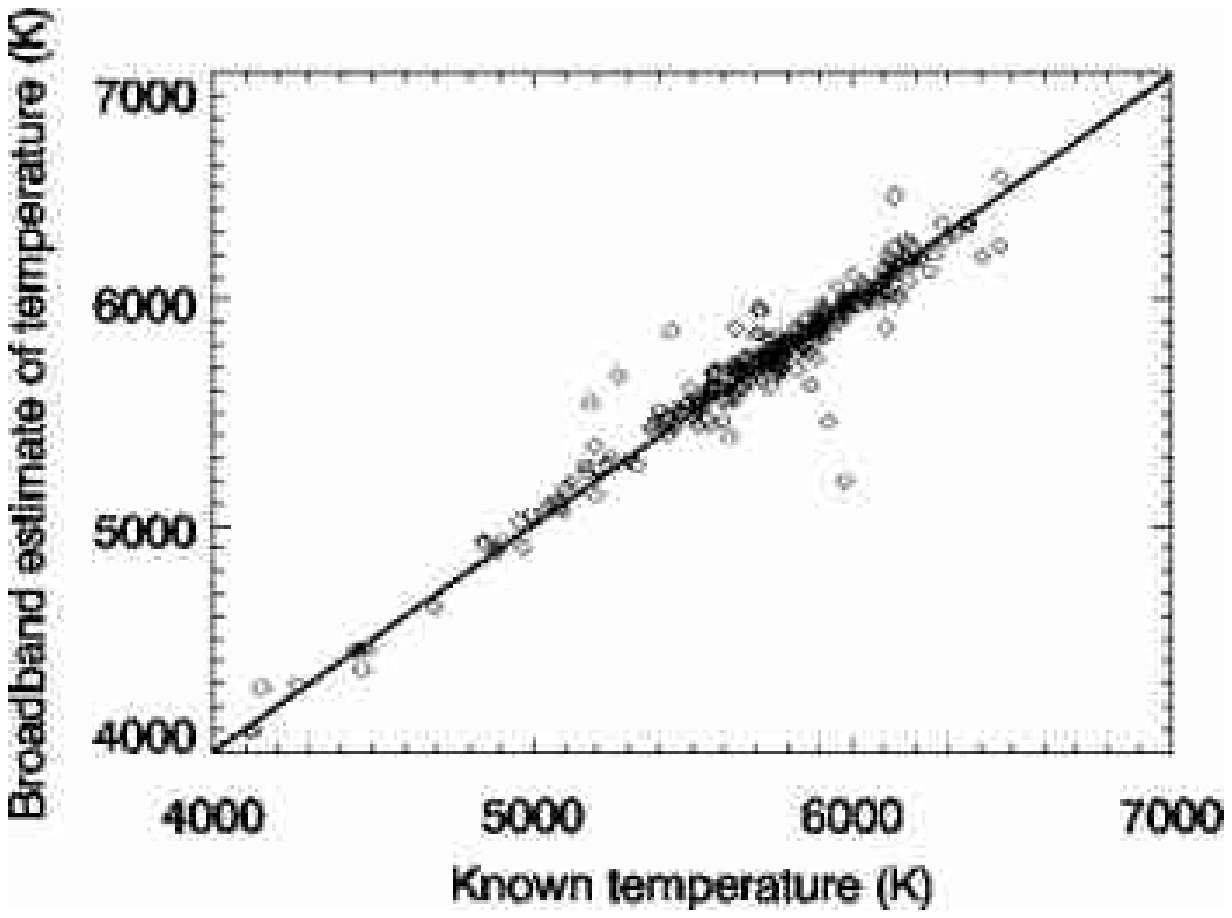}{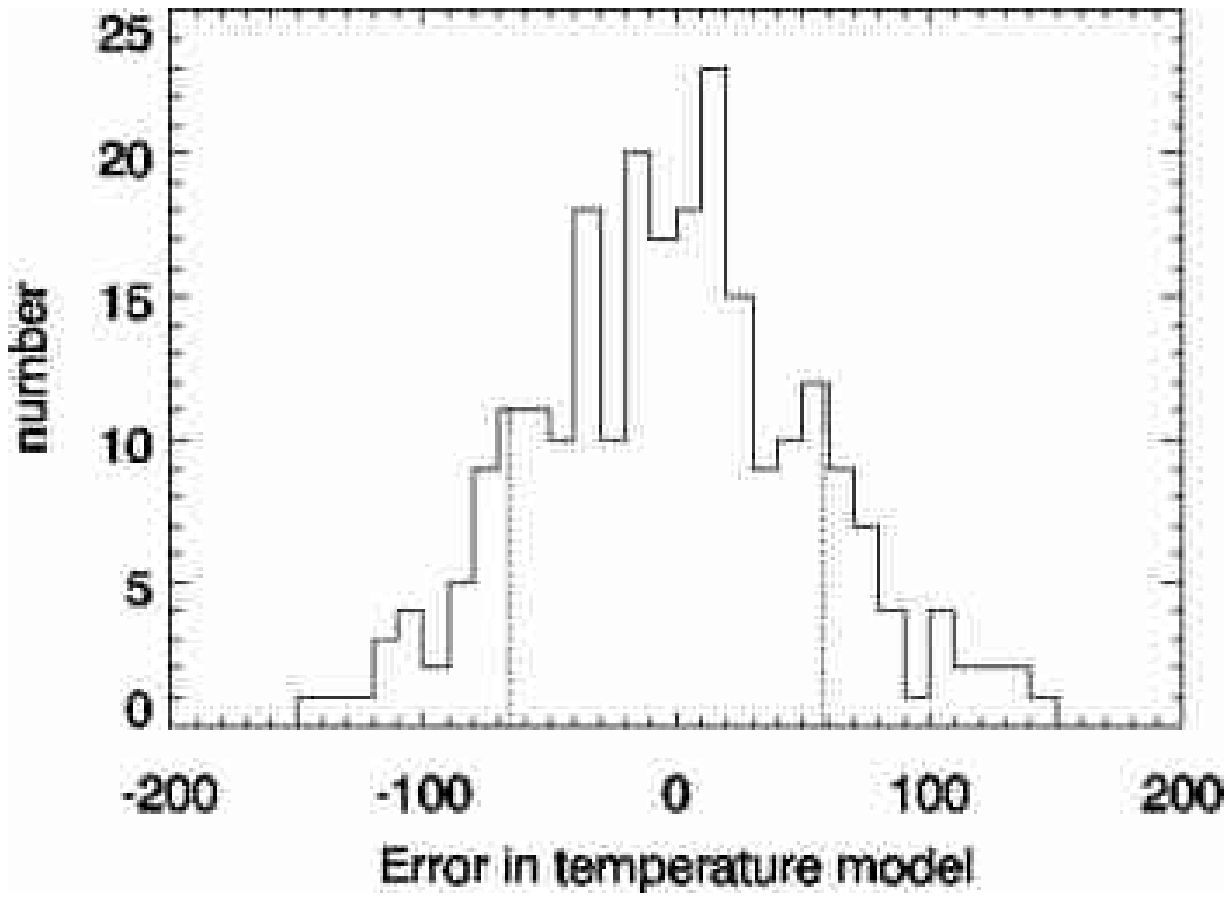}}
\caption{Upper left: Scatter plot of ``coarse'' \Teff\ model for 2433 Hipparcos stars.  Upper right: Histogram of residuals for the coarse \Teff\ model.  Bottom left: Scatter plot of ``fine'' \Teff\ model for 852 Valenti \& Fischer (2005) stars.  Bottom right: Histogram of residuals for the fine \Teff\ model.  The dotted vertical lines denote the one-sigma error intervals.}\label{fig1}
\end{figure}

\begin{table}
\begin{center}
\caption{Model parameters and errors for training sets and verification sets.\label{tbl-1}}
\begin{tabular}[t]{llllll}
\tableline\tableline
\multicolumn{1}{l}{} & \multicolumn{1}{l}{\Teff\ -coarse (K)} & \multicolumn{1}{l}{\Teff\ -fine (K)} & \multicolumn{1}{l}{distance (pc)} & \multicolumn{1}{l}{\FeH\ (dex)}  & \multicolumn{1}{l}{binarity} \\
\tableline
$N_{train}$\tablenotemark{a} &817 &233 & 751& 281& 2316\\
$N_{valid}$\tablenotemark{a} &860 & 233&731 & 307& 2358\\
$N_{test}$\tablenotemark{a} & 756& 262&782 & 262& 2326\\
$N_{bins}$\tablenotemark{b} & 1& 1& 6& 8& 11\\
Minimizer \tablenotemark{c}& $\sqrt{\chi^2}$ &$\sqrt{\chi^2}$ & $\sqrt{\chi^2}$&  $\sqrt{\chi^2}$&-\tablenotemark{h} \\
order\tablenotemark{d} & 2&2 &1 &1 &2 \\
$68\%+$\tablenotemark{e}  &125.4 &58.61 & (1.305)d& 0.123& - \tablenotemark{i} \\
$68\%-$\tablenotemark{e} &-211.6 &-65.81 & (-0.3866)d&-0.142 & - \tablenotemark{i}\\
$95\%+$\tablenotemark{f} &1229 & 216.7& (4.849)d&0.352 & - \tablenotemark{i}\\
$95\%-$\tablenotemark{f} &-2200 & -241.8& (-0.6716)d&-0.480 & - \tablenotemark{i} \\
output range\tablenotemark{g} &[3000,10000] & [4000,7000]& [0,300]&[-1.6,0.7]& - \tablenotemark{i}\\
\tableline
\end{tabular}

\tablenotetext{a}{The numbers of stars in each of the three-way data split (see \S 2.2) sets}
\tablenotetext{b}{Number of temperature bins used to divide the training set before computing the spline model (see \S 2.3)}
\tablenotetext{c}{Figure of merit used by the Levenberg-Marquardt minimizer to optimize the model (see \S 2.2)}
\tablenotetext{d}{Order of the polynomials in each of the temperature bins}
\tablenotetext{e}{Asymmetric one-sigma errors for model (positive and negative intervals), calculated as explained in \S 5}
\tablenotetext{f}{Asymmetric two-sigma errors for model (positive and negative intervals), calculated in the same way as the one-sigma errors}
\tablenotetext{g}{Range of model output over which the errors have been calculated for the test set}
\tablenotetext{h}{The binarity model used an alternate figure of merit (see \S 3.4.2)}
\tablenotetext{i}{The errors in the binarity estimate have been characterized by quoting a probability of multiplicity for individual stars, so the error intervals and output range are omitted.}
\end{center}
\end{table}

\subsection{Metallicity}

Approximate metallicities may be obtained from broadband data, given an accurate training set and a broad wavelength baseline for the photometry.  A great deal of heavy metal absorption occurs at short wavelengths, redistributing light to the red; proper motion also assists in differentiating the lowest metallicity halo stars from the local disk component.  Colors that include both optical and IR fluxes largely serve as temperature indicators, preventing spectral type contamination in the temperature bins.  These colors provide wide wavelength baselines that effectively break the degeneracy between temperature and \FeH.  As a training set, we use SME \FeH\ results from the Valenti \& Fischer (2005) catalog of over 1000 F, G, and K dwarfs with uncertainties of 0.03 dex.  Since the training set metallicities were obtained from spectra taken of single stars, it is required that the Tycho/2MASS fluxes also be of single stars.  Thus, any stars whose fluxes were likely to be the sum of multiple component stars, according to the Catalogue of Components of Doubles and Multiples (CCDM, Dommanget and Nys 1994) or Nordstrom et al. (2004), are not included in the training set.

The scatter plot and error residual histogram are shown in figure 2.  We have separately attempted \FeH\ polynomials on K giants, encountering abnormally large scatter due to the scarcity of stars in the training set.  This situation should improve as more spectroscopic \FeH\ observations for K giants become available.

\begin{figure}[htb]
\centerline{\plottwo{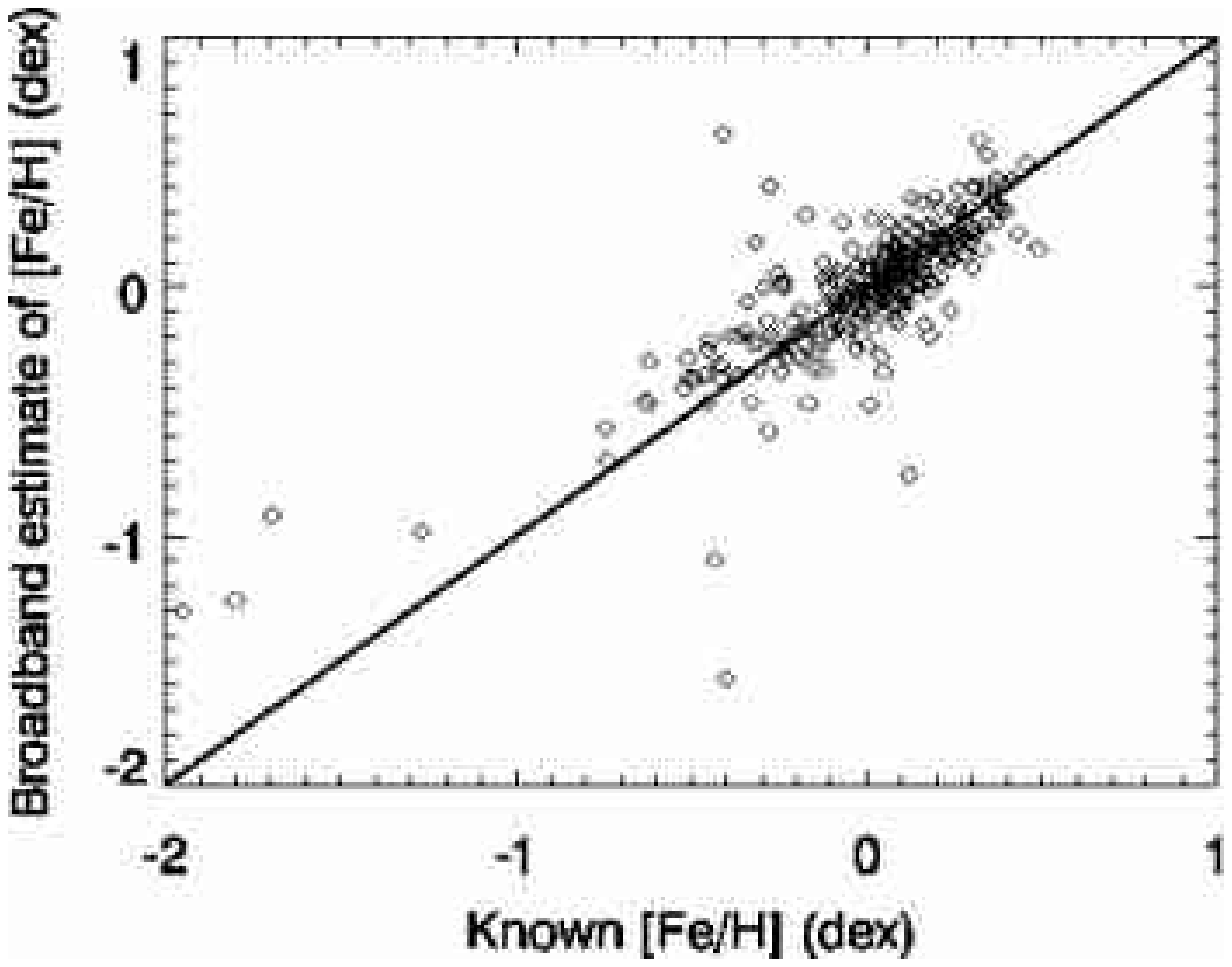}{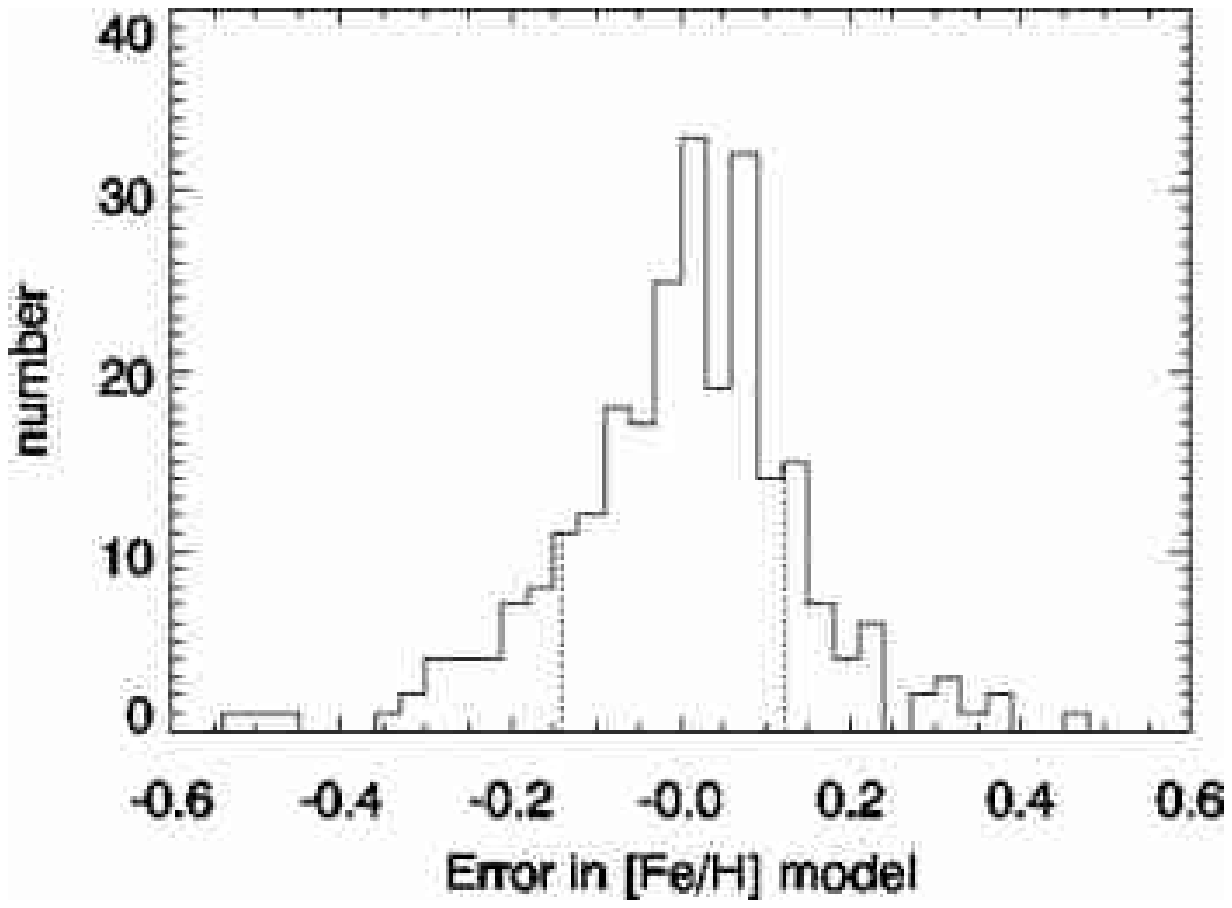}}
\caption{Left: Scatter plot for \FeH\ polynomials, including dwarfs of all temperatures.  Right: Histogram of residuals for \FeH\ fit.  Units are in decades.  The dotted vertical lines denote the one-sigma error intervals.}\label{fig2}
\end{figure}

\subsection{Distance}

Proper motion may serve as a proxy for distance when no trigonometric parallax is available.  We have attempted to approximate distance explicitly from proper motion and colors alone by fitting to Hipparcos parallax first and then converting to distance.  The model scatter plot and error histogram are shown in figure 3, and the model parameters are given in Table 1.  The errors are given as a function of distance output by the model.  The approach is similar to that of the reduced proper motion technique (see, e.g., Luyten 1922, Chiu 1980, Gould and Morgan 2003), although it includes the effects of reddening.  A very large pool of Hipparcos stars is available for the training set, making possible stricter inclusion cuts.  We remove any known variables or multiple stars.  In addition, we only consider stars that have temperatures recorded in Cayrel de Strobel (1997) or Valenti \& Fischer (2005) so that we may use the spline formulation outlined in \S 2.  The results are comparable to those of reduced proper motion (Gould and Morgan 2003) with reddening included, although several observational biases enter into account.  For example, the fact that the redder giants have greater distances misleads the polynomial's interpretation of color.

\begin{figure}[htb]
\centerline{\plottwo{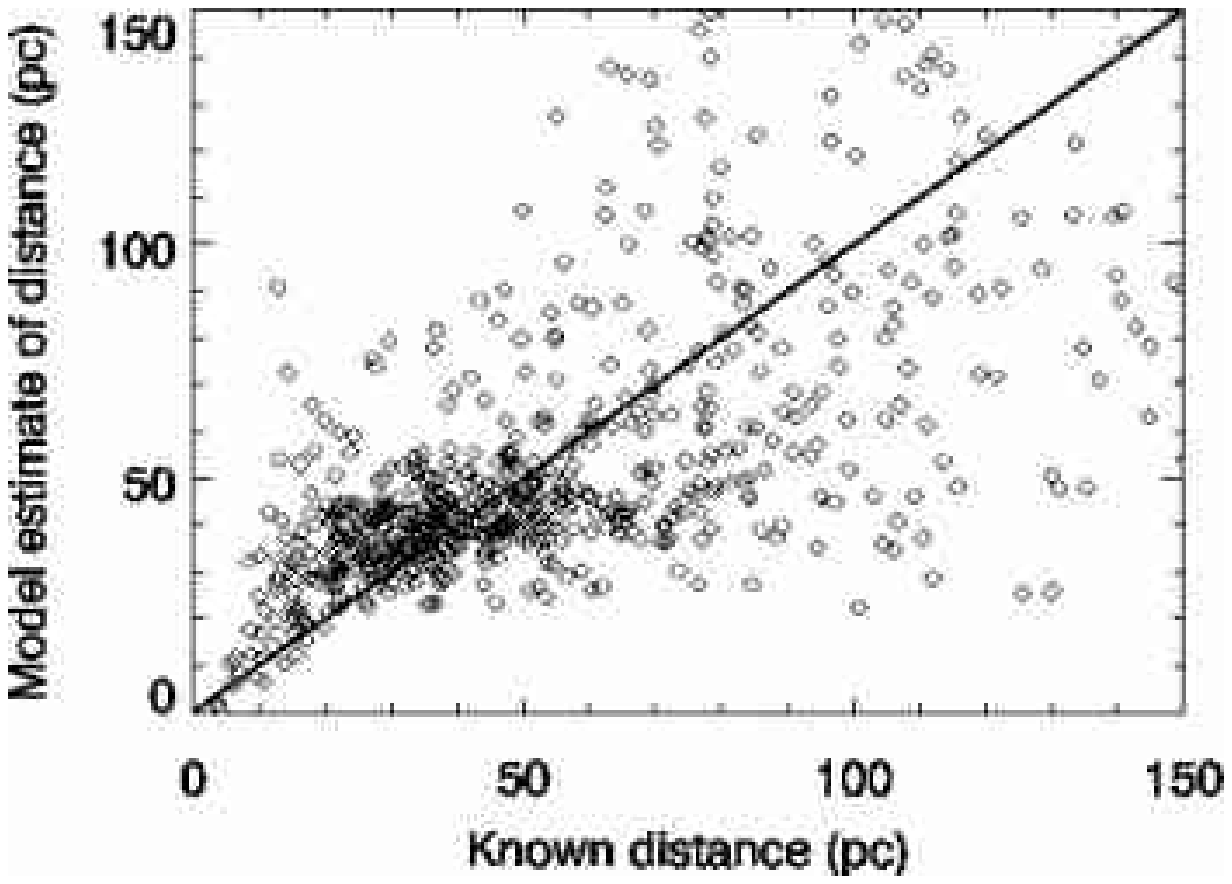}{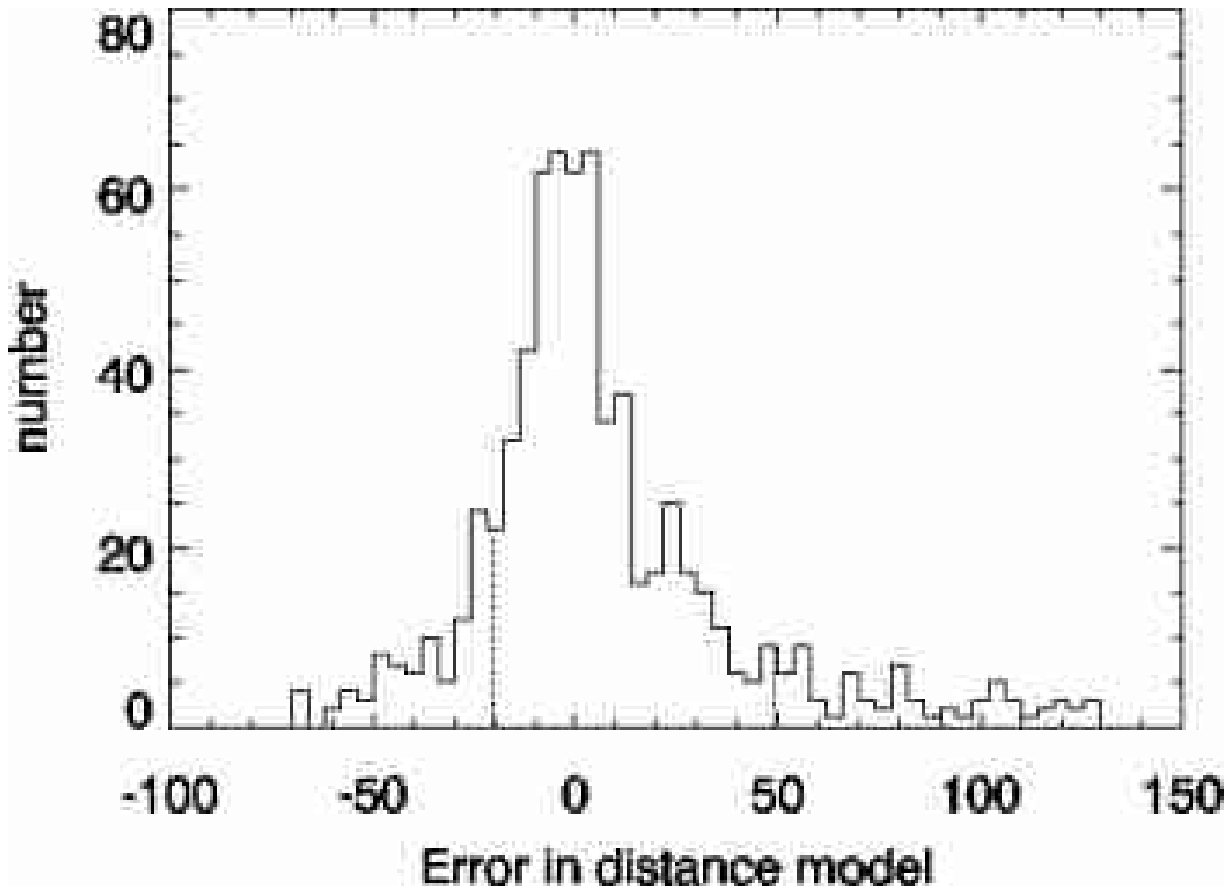}}
\caption{Left: Scatter plot for distance polynomial.  Units are in parsecs.  Right: Histogram of residuals for distance polynomial.  The dotted vertical lines denote the one-sigma error intervals.}\label{fig3}
\end{figure}

\subsection{Binarity}

\subsubsection{The Model}
Binaries with intermediate mass ratios should be identifiable by optical and IR colors alone, as their integrated photometry is a composite of two blackbody SED's that peak at separate wavelengths.  We have produced a ``binarity'' model that, through this effect, attempts to identify binaries within a certain mass ratio range.  Several applications can benefit from the removal of binaries from target lists, most notably the radial velocity surveys that are blind to planets around spectroscopic binaries.  We are most interested, however, in preventing doubles from corrupting the \FeH\ estimates.  Binarity is here quantified by assigning a value of $1$ to known doubles and a value of $0$ to both singles and undetectable (via photometry) doubles.  Binaries with mass ratio near unity $(\mrat\ < 1.25)$ are indistinguishable from single stars with color information only.  Binaries with large mass ratios $(\mrat\ > 3)$ are also similarly difficult to flag because the secondary produces little flux compared to the primary.  Thus we only focus on finding binaries whose absolute $V$ magnitudes differ by more than 1 magnitude and whose $Ks$ magnitudes differ by less than 3 magnitudes.  These criteria have been chosen to ensure that the $V-K$ difference between the components exceeds the typical color error in our Tycho/2MASS overlap list.

The training parameter is discrete (0 or 1) in the binarity model, but the model output is continuous.  Instead of a scatter plot and error histogram for this model, which do not contain helpful visual information, we display plots of pass rates for two models in figure 4.  This plot shows the percentage of doubles above a certain binarity threshold (the threshold ranging from -0.5 to 1.0) as a function of the percentage of singles below the threshold.  A perfect single/double discriminator would appear as a horizontal line at $100\%$.  A ``random pick'' discriminator would appear as a diagonal line with a slope of $-1$.  Essentially, the binarity model shown here can be used to rank a target list and isolate groups of stars that have a lower likelihood of containing doubles that satisfy $1.25 <$ \mrat\ $< 3.0$.  For any sample size given by the percentage on the x-axis, the reduction in the number of doubles is given by the percentage on the y-axis.  The number of these types of detectable doubles is small in a magnitude-limited sample, so the savings are not necessarily large.

\begin{figure}[htb]
\centerline{\plottwo{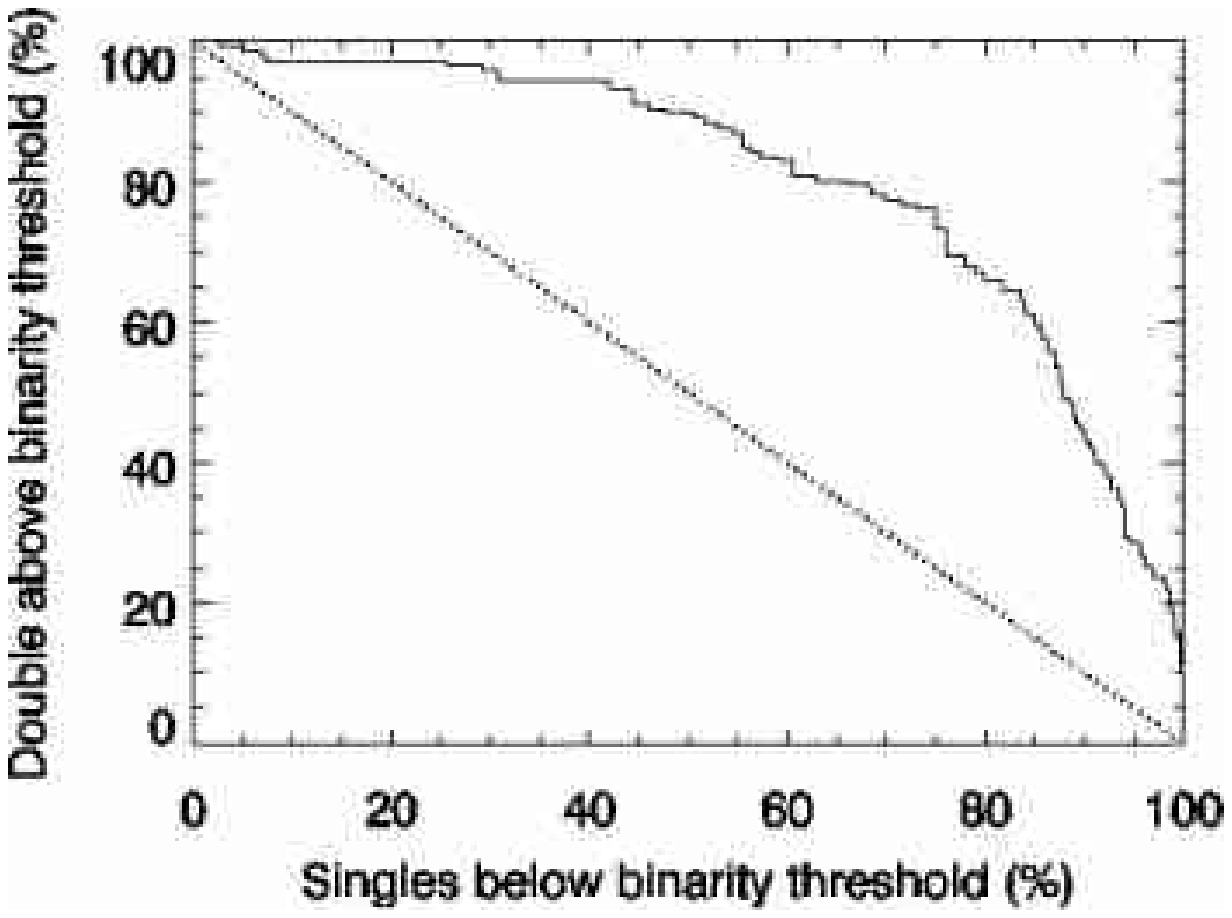}{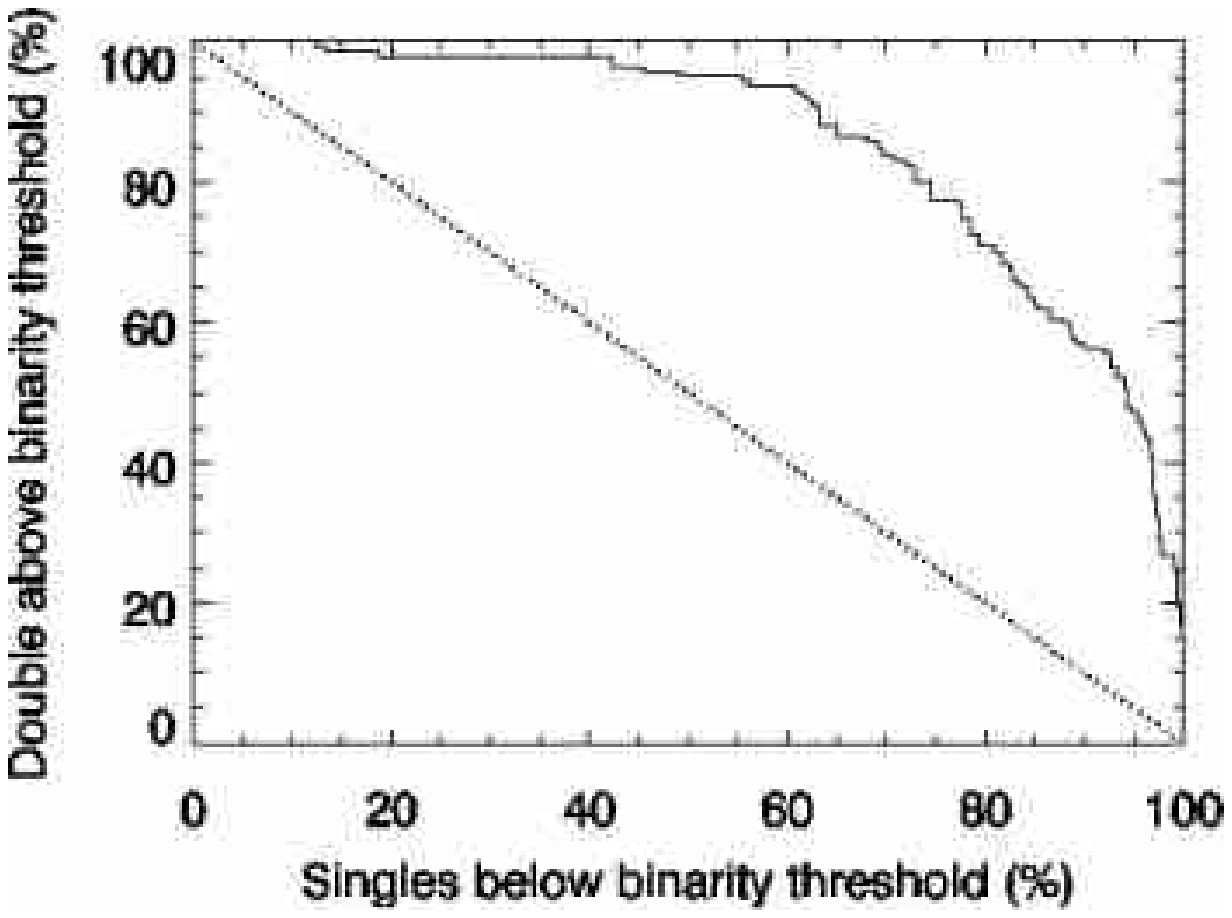}}
\caption{Pass rate plot for two binarity models.  Each point represents a potential cut of the training set population to eliminate as many doubles as possible.  The x-value represents the percentage of singles that would remain in the sample for a given cut and the y-value represents the percentage of detectable doubles that would be eliminated.  Note that half of the detectable doubles can be removed from the target list without affecting the total size appreciably.  Left:  Kroupa, Tout, \& Gilmore (1993) IMF training set.  Right:  Same IMF, but $\alpha = -2.35$ for $M > 1 \Msun\ $.}\label{fig4}
\end{figure}

\subsubsection{Simulating a magnitude-limited population}
We have chosen to simulate the binarity training set rather than use an existing set whose stars have known multiplicity.  A binarity training set must be composed of stars for which the multiplicity and mass ratio are known, so that these values can be connected to photometry.  To reduce error in applying the model, the frequency distributions of all relevant parameters (metallicity, mass ratio, etc.) must match the application set as closely as possible.  Unfortunately, all known sets that satisfy the first requirement (e.g., Duquennoy and Mayor 1991, Dommanget and Nys 1994, Nordstrom et al. 2004, Setiawan et al. 2004, etc.) do not satisfy the second when the application set is Tycho 2, a magnitude-limited survey complete to $m_v = 11.5.$  We satisfy the second constraint by drawing samples of stars with known binarities in a highly biased fashion and attempt to remove these biases by populating the simulated training set according to the spectral type distributions of a magnitude-limited population.  We manually simulate double systems by summing the fluxes of single stars stochastically.  We do not use giants as companions because the enormous flux would swamp that of the secondary.  

To create the training set, we begin with a population of stars close and bright enough to ensure that most multiples have been marked by other studies.  We include all F, G, and K dwarfs with distance modulus less than 2.5 and trigonometric distance less than 60 pc.  All known M dwarfs in Hipparcos are included, to be used as a pool of secondaries.  These FGKM stars are the most likely of all the Hipparcos stars to have their multiplicity correctly recorded in either the Catalogue of Components of Doubles and Multiples (CCDM) or the Nordstrom et al. (2004) radial velocity survey of F and G dwarfs.  Stars flagged as doubles or multiples are removed from the set.

To simulate a Tycho 2 pool, we assume that stars have companions with a mass-dependent frequency proportional to the IMF.  We ignore the overabundance of binaries with unity mass ratio.  We utilize a Kroupa, Tout, and Gilmore (1993) IMF, with a steep ($\alpha = -2.7$ for $M > M_{\odot}$) dropoff at high masses and shallower slopes at intermediate and low mass ($\alpha = -2.2$ for $0.5 M_{\odot} < M < M_{\odot}$ and $\alpha = -1.2$ for $M < 0.5 M_{\odot}$).  We then randomly pair stars according to the IMF, calibrating the total number of single stars versus the number of multiples by assuming that $70\%$ of all G dwarfs have secondaries more massive than $0.1 M_{\odot}$ (Duquennoy and Mayor 1991).  Any pairs whose absolute magnitudes differ by less than 1 magnitude in V or by more than three magnitudes in Ks are labeled as single and assigned a binarity of zero.  Assuming a constant star formation rate in the disk over 10 Gyr, we remove stars from the simulated samples according to their probability of leaving the main sequence before reaching the present time.  We also remove stars according to the probability function $1-P\;\propto\;10^{-\frac{3}{5}M_v},$ the probability of observation in a magnitude-limited survey. 

After trimming the simulated sample in the manner above, a population remains whose numbers peak at a temperature of $\sim5200\;K$.  Roughly $7\%$ are detectable doubles satisfying the mass ratio criteria specified above.  We do not optimize the binarity model with a $\chi^2$ minimization, as for the other models; instead, we are more interested in making cuts in target lists and maximizing the number of doubles eliminated when doing so.  We calculate a new figure of merit that captures this interest.  For any model, we can compute the binarity value (ranging from zero to one) for all simulated stars.  We find the binarity threshold that divides the set of true singles into two sets of equal size.  The figure of merit we use is the percentage of detectable doubles above this binarity threshold.  A perfect discriminator would have a figure of merit of $100\%$; a ``random-pick'' discriminator would have a figure of merit of $50\%$.  The final model chosen has a figure of merit of $89.8\%$.  We have also generated a second binarity model using a modified training set, one with a shallower Salpeter IMF at high masses ($\alpha = -2.35$ for $M > M_{\odot}$).  This second model has a final figure of merit of $95.1\%.$  We apply both of these models to Tycho 2 for comparison (see figure 4 or 8 for comparisons).  

For the Tycho 2 stars to which we intend to apply these models, the value output from the binarity model is less useful than the actual probability of a star being double or multiple.  This important parameter is equal to the ratio between the number of labeled doubles to the total numbers of stars in a single binarity bin in the training set.  These probabilities are accurate as long as the relative proportions of multiples to single stars is nearly correct in the training set.  The data set we publish includes the calculated probability of multiplicity for each star (from both models referred to above) as well as the estimated error on these values.  These errors are estimated from the photometry error only, and not the intrinsic scatter errors in the models, as these latter errors determine the probability of multiplicity itself (i.e., scatter error is the only reason that the probabilities are not $0\%$ or $100\%$ exactly for all stars).  In practice, these errors will be dominated by the lack of similarity between the simulated training set and the observed Tycho 2 set, which we do not attempt to quantify.

Again, the probabilities of multiplicity given for each star ignore doubles with similar masses.  Photometry is practically blind to close binaries of this type.  In this paper, we refer to the ``probability of multiplicity'' as the probability of finding a double that falls within only a small range of mass ratios $(1.25 < \mrat\ < 3)$.  In addition, it should be noted that the binarity and the probability of multiplicity are largely meaningless for giants and subgiants, as the training set only includes dwarfs.  Giants have been ignored here because (1) giant/dwarf pairs would be invisible to photometry due to the considerable difference in absolute magnitude, (2) giant/giant pairs are more rare than dwarf doubles, (3) radial velocity measurements are difficult to obtain for giants, and (4) giants usually engulf companion stars.

\section{APPLICATION TO TYCHO}

\subsection{Model Output}
We present catalogs containing the results of applying these model polynomials to 2,399,867 stars from the Tycho 2 data.  This subset consists of stars whose photometry, proper motions, and coordinates exist in both Tycho and 2MASS and whose 2MASS equivalents were not ambiguously matched to multiple Tycho 2 stars.  A total of 140,226 stars were excluded on these grounds.  For the majority, estimated \Teff and distance are given with errors.  We provide \FeH\ and the probability of multiplicity for the stars that appear to be dwarfs (see \S 4.2).  We adopt the following procedure for computing model values:  (1) Estimate distance and effective temperature using coarse polynomial (see 3.1); (2) Isolate dwarf pool using colors and distance information; (3) Re-estimate temperature with fine polynomial, estimate reddening, and remove possible contaminants with this new information; (4) Calculate \FeH\ and binarity for dwarfs; and (5) Calculate scatter error and photometry error for all parameters.  The procedure for isolating dwarfs is given in \S 4.2.

The error intervals are derived from the residuals and are given as functions of the output of the polynomials.  Photometry errors are included as described in \S 5.  Large errors occur for stars with mismatched Tycho/2MASS photometry or stars with low-quality measurements.  Histograms of the polynomial outputs are shown in figure 5 for \Teff\ and distance, which are estimated for all 2,399,867 stars.  The dashed histograms have been made for the stars in both Hipparcos and the Tycho 2 set.  The solid histograms are for all Tycho 2 stars in the dwarf pool.

\begin{figure}[htb]
\centerline{\plottwo{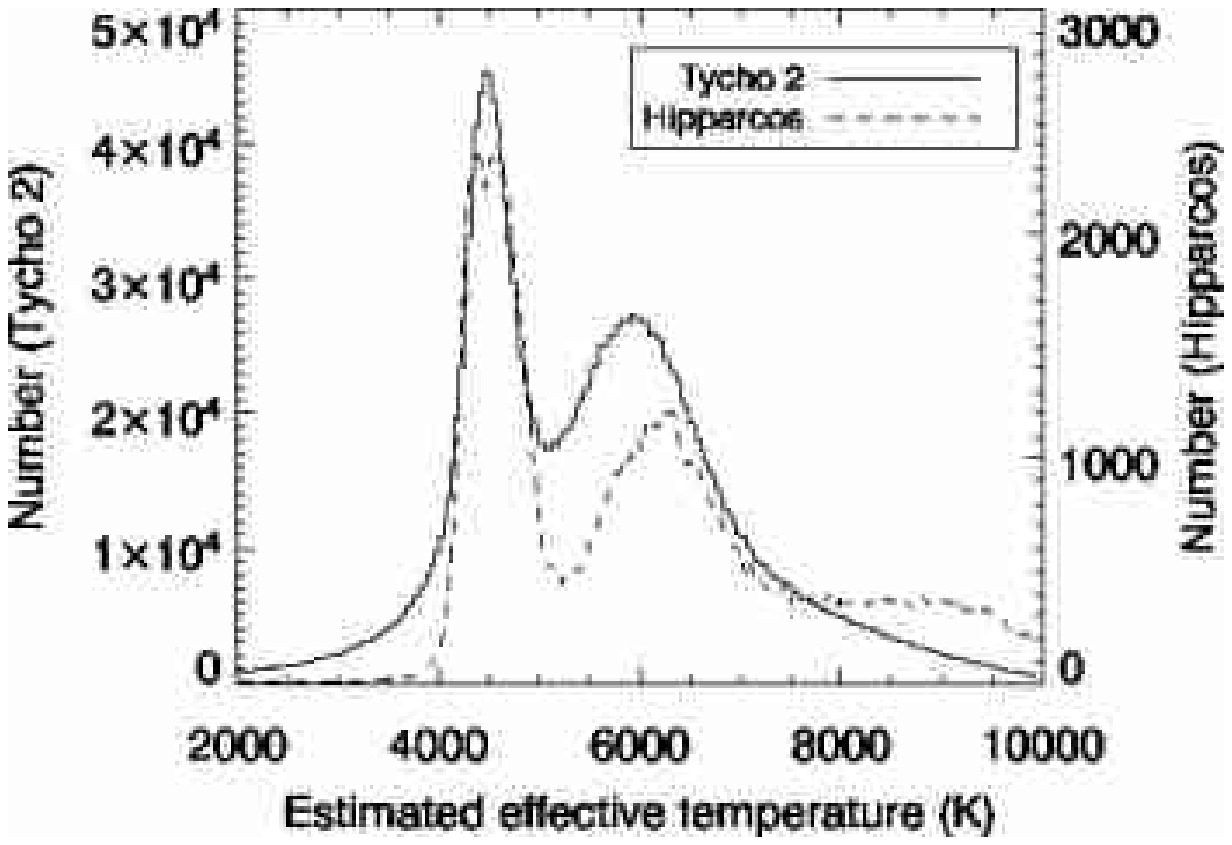}{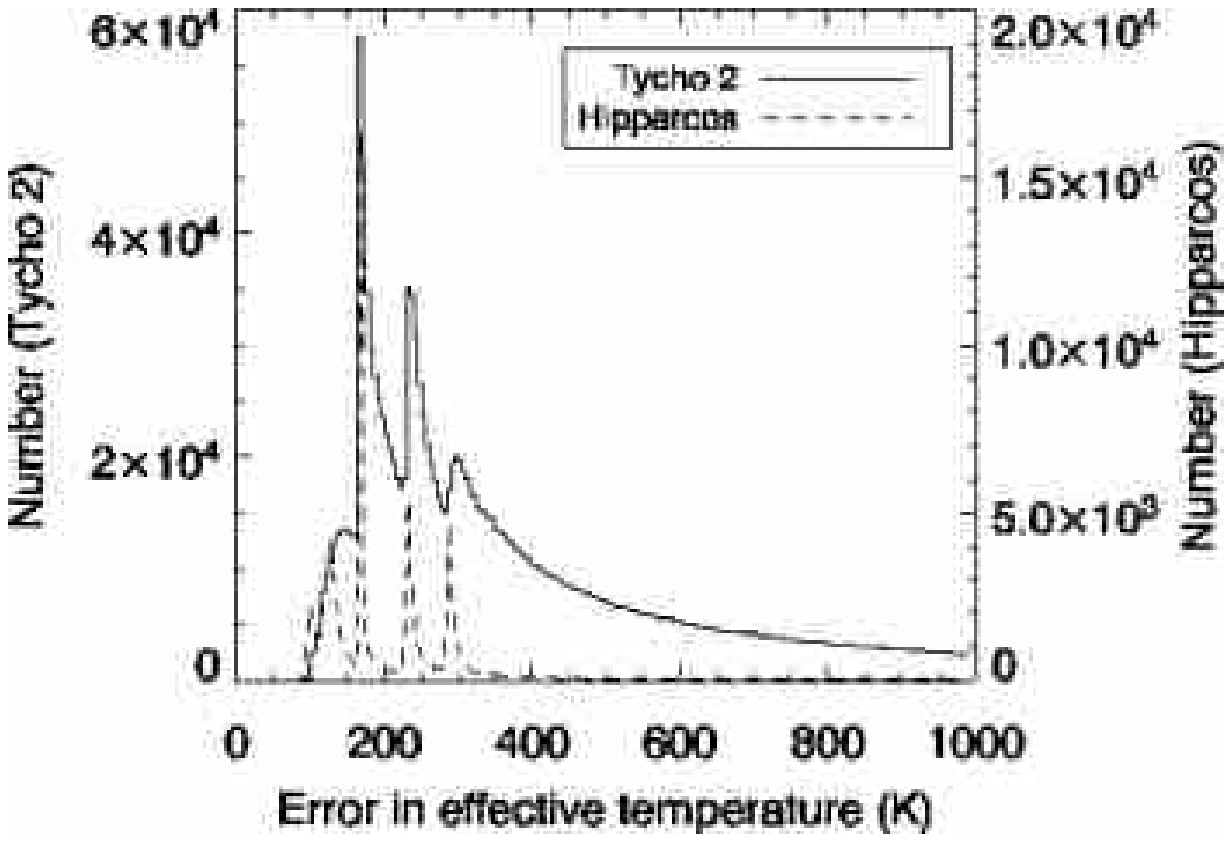}}
\centerline{\plottwo{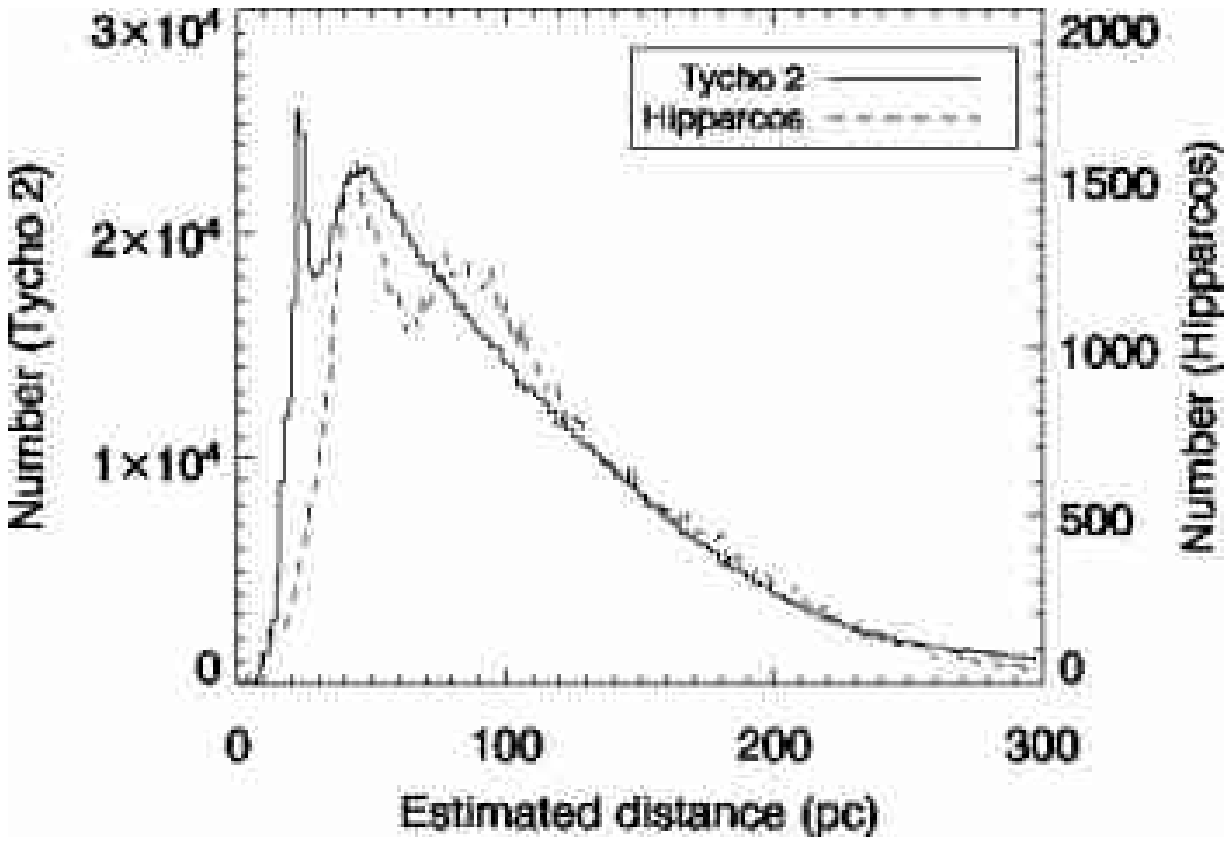}{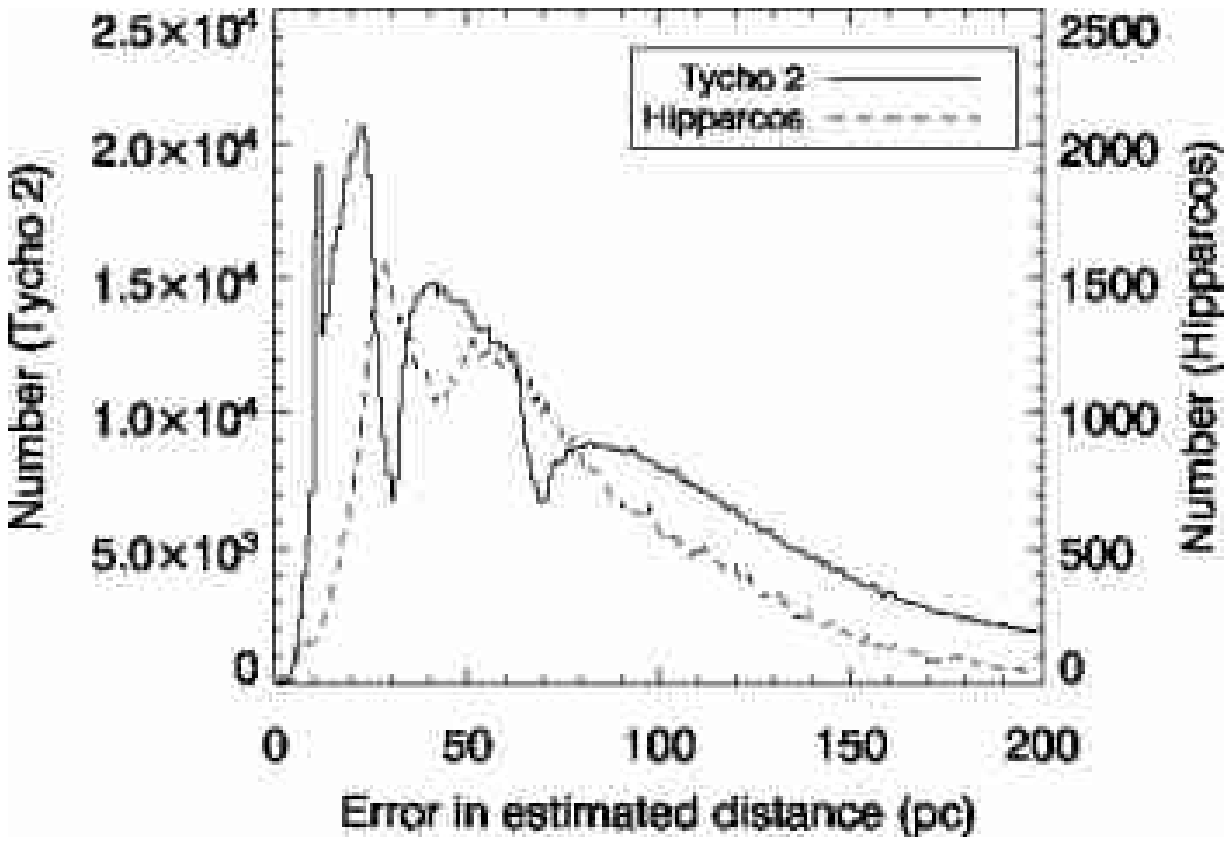}}
\caption{Top left: Histogram of coarse \Teff\ model output for the Tycho 2 set, consisting of 2,399,867 stars.  Output for the Hipparcos subset of this set is shown as a dashed
histogram for all plots.  Top right: Histogram of errors of coarse \Teff\ model for all Tycho 2 stars.  Error widths are calculated by averaging the positive and negative
intervals and are representative of a one-sigma error.  Bottom left: Histogram of distance polynomial output for all Tycho 2 stars.  Bottom right:  Histogram of errors from the
distance model.  Error widths are calculated by averaging the positive and negative intervals.}\label{fig5}
\end{figure}

\subsection{An expanded sample of Dwarf Stars in the Solar Neighborhood}
The model outputs were used to define a dwarf pool using the cutoffs
$$d_{est} < 200\; pc, \:\:\:\:\:\:\:\:3850\; K < T_{est} < 7200\; K, \:\:\:\:\:\:\:\: M_v < 8 - \frac{8(T_{est} - 3850\; K)}{3350\;K}$$
where the absolute magnitude is determined from the estimated distance.  The last criterion defines a discriminatory line in the HR diagram that eliminates giants from the pool based on distance and absolute magnitude information.  We also eliminate stars with $V-K > 4$ to lower the giant contamination and eliminate stars with large distance or temperature errors ($\sigma_d > 0.7d$ or $\sigma_T > 0.1T$).  After defining this pool, we estimate the reddening with the Schlegel et al. (1998) dust maps in the following way.  The dust maps provide E(B-V) for a line of sight exiting the galaxy for all galactic coordinates.  We assume that the galactic dust is distributed in a double exponential form, with the density falling off with disk height using a scale factor of 350 pc and with Galactic radius using a scale factor of 3000 pc.  This model sets the sun at 8.5 kpc from the center at a height of 0 pc.  We also assume that the dust-to-gas ratio is uniform and constant and that E(B-V) is proportional to true distance along any line of sight.  To estimate the reddening for a star, we first calculate an expected absolute magnitude from a main sequence fit to V-K color, which gives a distance.  $E(B-V)$ is proportional to the line integral 
$$\int_{0}^{d} dL \:e^{-z / z_t} e^{-R/R_t}$$
where $d$ is the total distance to the object, $z$ is the height above the Galactic plane, $R$ is the radius from the Galactic center, $z_t = 350$ pc is the disk scale height, and $R_t = 3000$ pc is the disk scale radius.  We evaluate this integral for the estimated distance to the star and divide the result by the integral to $d = \infty.$  We then multiply this ratio by the Schlegel et al. extinction value $E(B-V)$ to arrive at an estimated $E(B-V)_{obj},$ the extinction to the star.  We then update the colors using the Rieke \& Lebofsky (1985) interstellar extinction law, with the foreknowledge that Tycho magnitudes roughly parallel the Johnson system.  We also update the star's expected absolute magnitude and distance, assuming $A_V \;=\; 3.1 E(B-V)_{obj},$ and repeat the entire process.  For the majority of stars, two or three iterations result in convergence, so we do not repeat beyond three iterations for any of the stars.  A minority of stars near the galactic plane have diverging extinctions for more than three iterations, but the colors of these stars are so reddened and uncertain that they are likely not worth pursuing for surveys.  In addition, we only keep the color correction if $E(B-V)_{obj} < 0.2$. 

We calculate improved effective temperatures, metallicities, and the probability of multiplicity for all 354,822 dwarfs remaining in the dwarf pool.  The new temperatures are calculated using the ``fine'' temperature model which is optimized for dwarfs.  Histograms of temperature and [Fe/H], with errors, and estimates of extinction $E(B-V)$ are shown in figure 6.  Histograms of probability of multiplicity, estimated via two different models, are shown with errors in figure 7.  Note that the output of the binarity models is discrete when expressed as probabilities that range between $0$ and $1.$  The error histograms tend to have a great deal of structure because the errors are a combination of discrete model error and continuous photometric error (see \S 5).  Note qualitatively that a large number of Tycho 2 stars that are not in Hipparcos retain good one-sigma errors for \FeH\ and $T_{\rm eff}$.

Figure 8 displays several comparisons between model output to test for biases in the dwarf pool, presented as star density plots.  The upper left plot shows the effective temperature for the dwarf pool, estimated from the fine temperature model, plotted against the calculated temperature error.  The positive and negative error bars are averaged for each star to give a single estimate of the one-sigma error.  The step-like structure in the plots is again due to the discrete model error.  The upper right plot shows the estimated metallicity \FeH\ against the \FeH\ error for the dwarf pool.  Unlike the previous temperature error plot, the mean estimated \FeH\ shifts when the photometry error increases.  This bias can be avoided by choosing a \FeH\ error cutoff when analyzing stars, e.g., $\sigma_{\FeH\ } < 0.3.$

Also displayed in the bottom left of figure 8 is a plot of estimated \FeH\ versus the probability of multiplicity for the KTG model.  As mentioned in \S 3.4, we constructed a binarity model to isolate doubles and prevent \FeH\ estimates from being corrupted, as we would expect both models to look for similar signals in the IR colors.  It is clear from this plot that the two models are indeed confounded; stars that are flagged as doubles are also more metal-poor, on average.  We have determined that these stars are likely true multiples with underestimated [Fe/H].  The \FeH\ model looks at $B-V$ exclusively for stars in a given temperature bin (i.e., stars in a certain $V-K$ range).  A blue decrement is a sign of metal-rich composition.  Adding a smaller secondary star to a primary SED increases $V-K,$ $V-H,$ and $J-K$ relative to $B-V$ and $B-J,$ which don't change significantly.  However, this places the star in a cooler temperature bin because the temperature models consider $V-K$ color primarily.  The pair thus has abnormally blue $B-V$ color for the cooler bin and is immediately assigned an underestimated [Fe/H].  This bias may be avoided by using the binarity models to eliminate likely doubles before analyzing [Fe/H].

The bottom right of figure 8 is a comparison between two different binarity models.  Notice that when the IMF is modified with a shallower Salpeter IMF at high masses, the probability of multiplicity is slightly overestimated relative to the unmodified Kroupa, Tout, \& Gilmore (1993) IMF.  We suspect that this is due more to variation in model parameters than the modification in the IMF.  The model parameters vary widely between these two binarity models, particularly in the number of temperature bins chosen (22 for the modified IMF, 11 for the unmodified IMF).  However, the probability of multiplicity is a very approximate estimate, and this discrepancy is within the errors for both models.  

\begin{figure}[htb]
\centerline{\plottwo{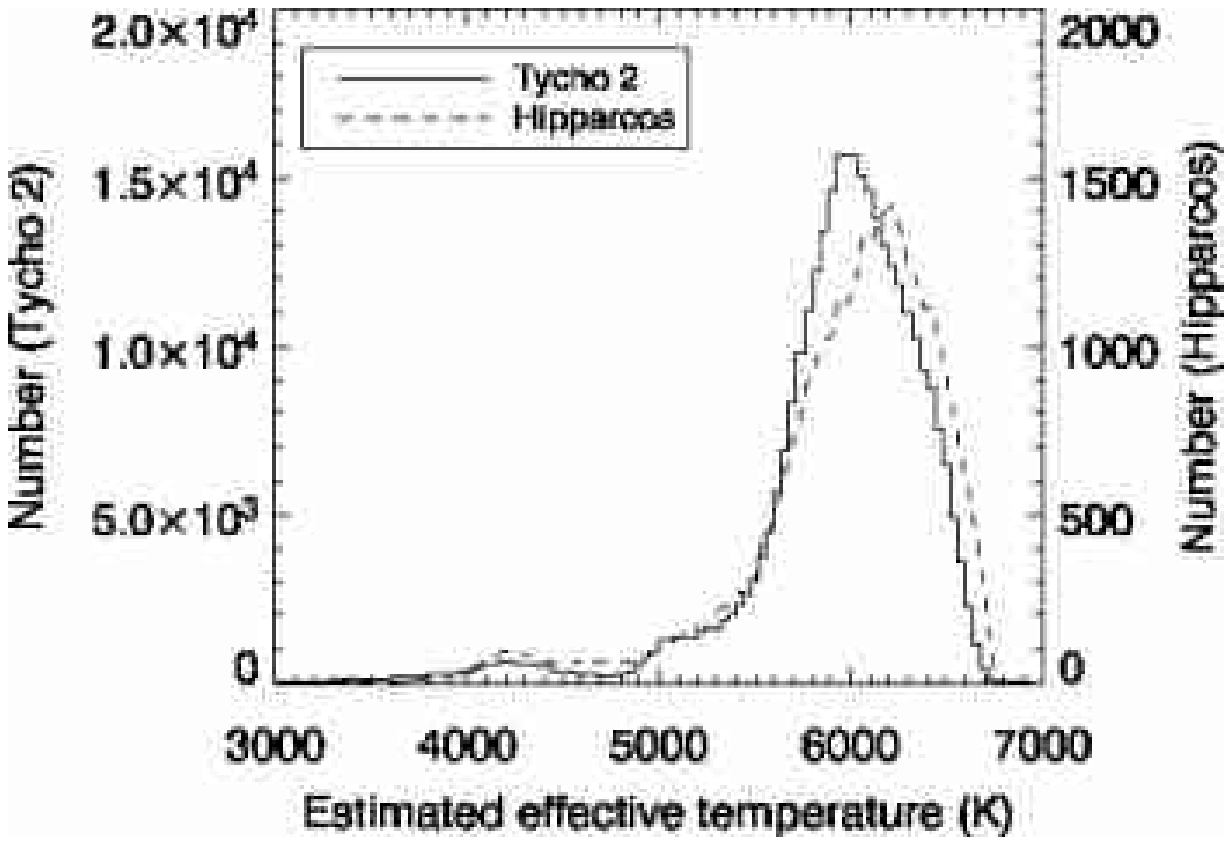}{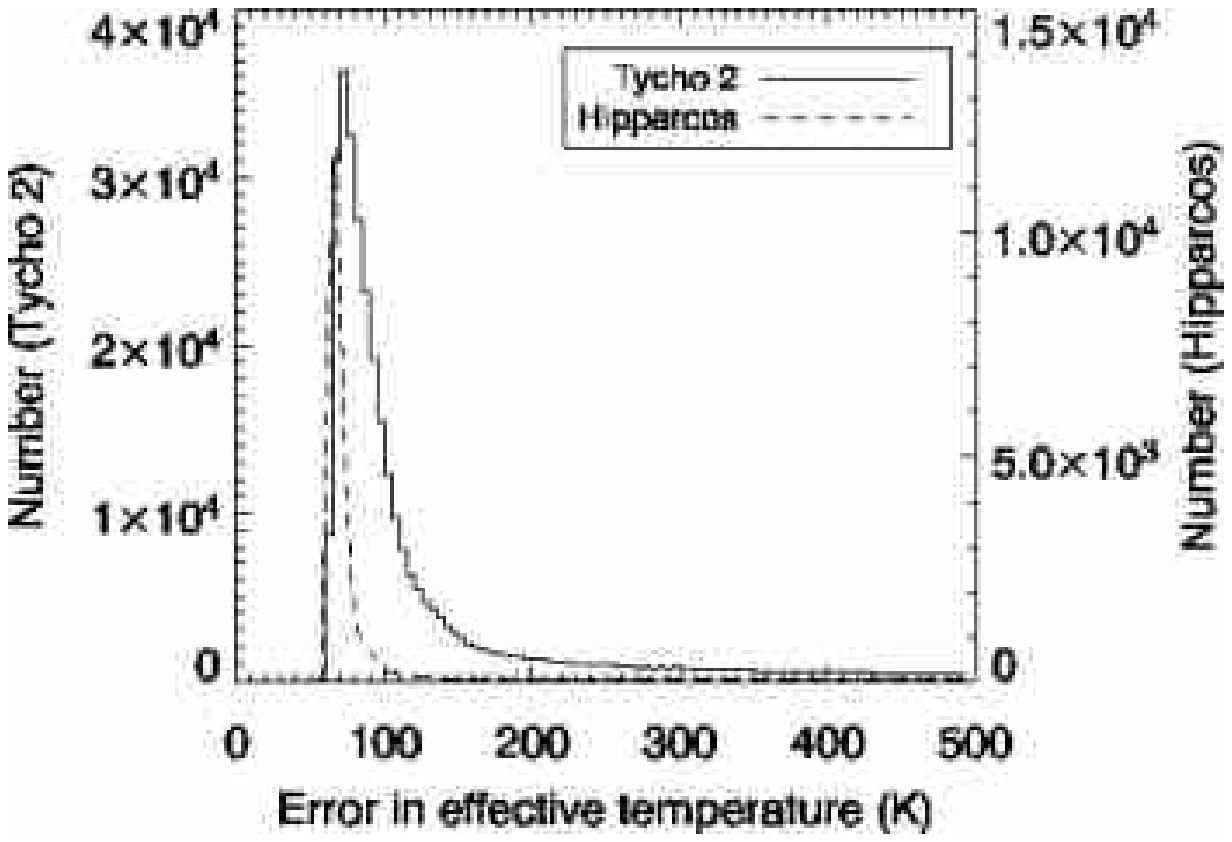}}
\centerline{\plottwo{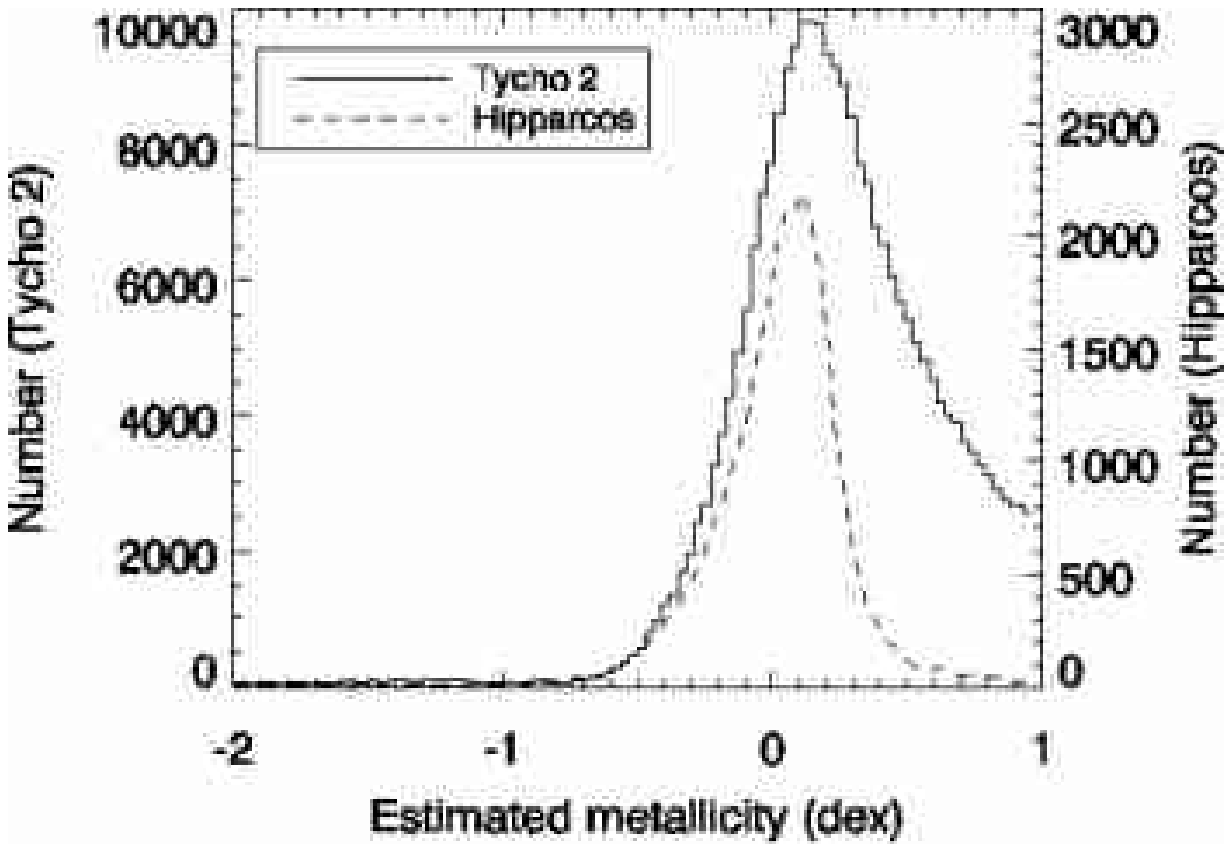}{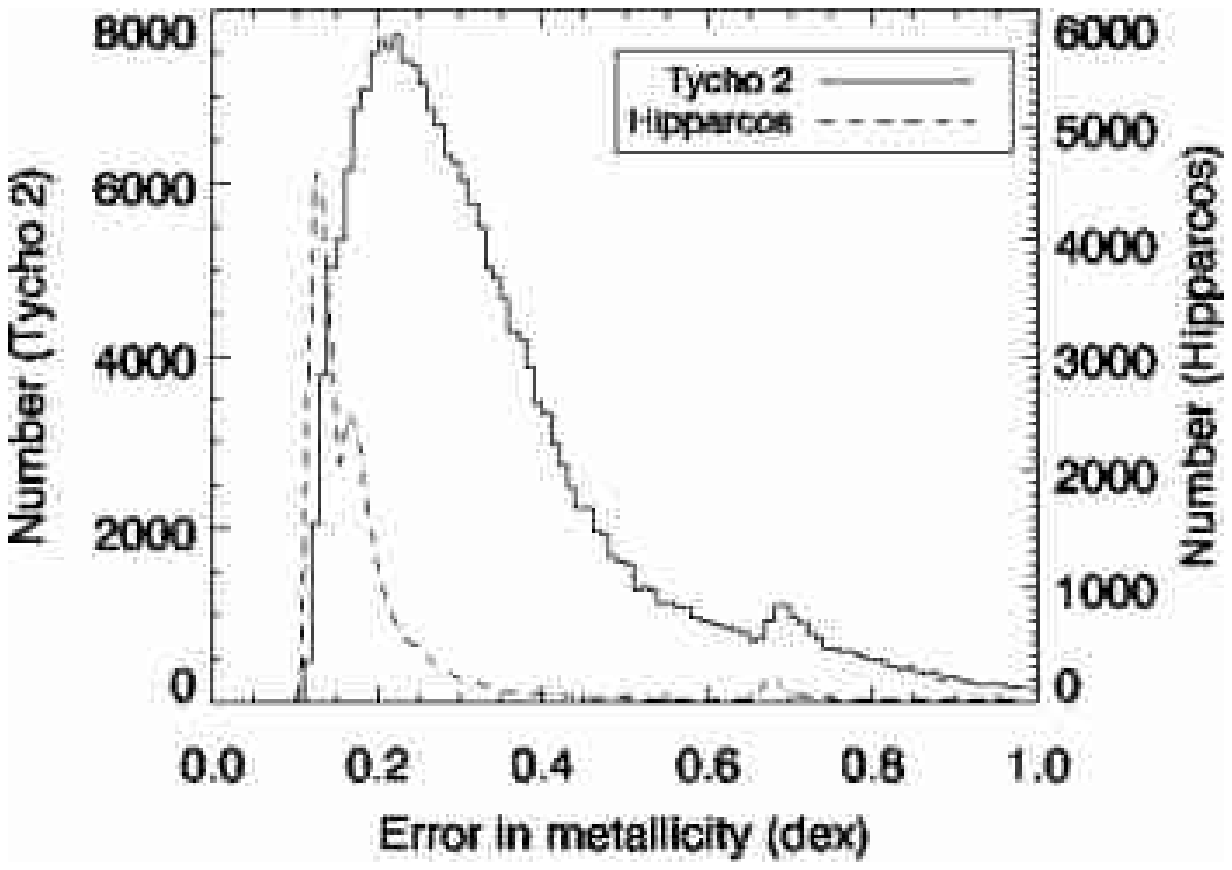}}
\caption{The solid lines are histograms for Tycho 2 stars in the dwarf pool; the dashed lines are histograms for the Hipparcos stars in the pool.  Top left:  Histogram of fine
temperature estimates for dwarf pool.  Top right:  Histogram of fine temperature errors for the dwarf pool.  Error widths are calculated by averaging the positive and negative
intervals and are representative of a one-sigma error.  Bottom left:  Histogram of \FeH\ model output.  Bottom right:  Histogram of \FeH\ errors.}\label{fig6}
\end{figure}

\begin{figure}[htb]
\centerline{\plottwo{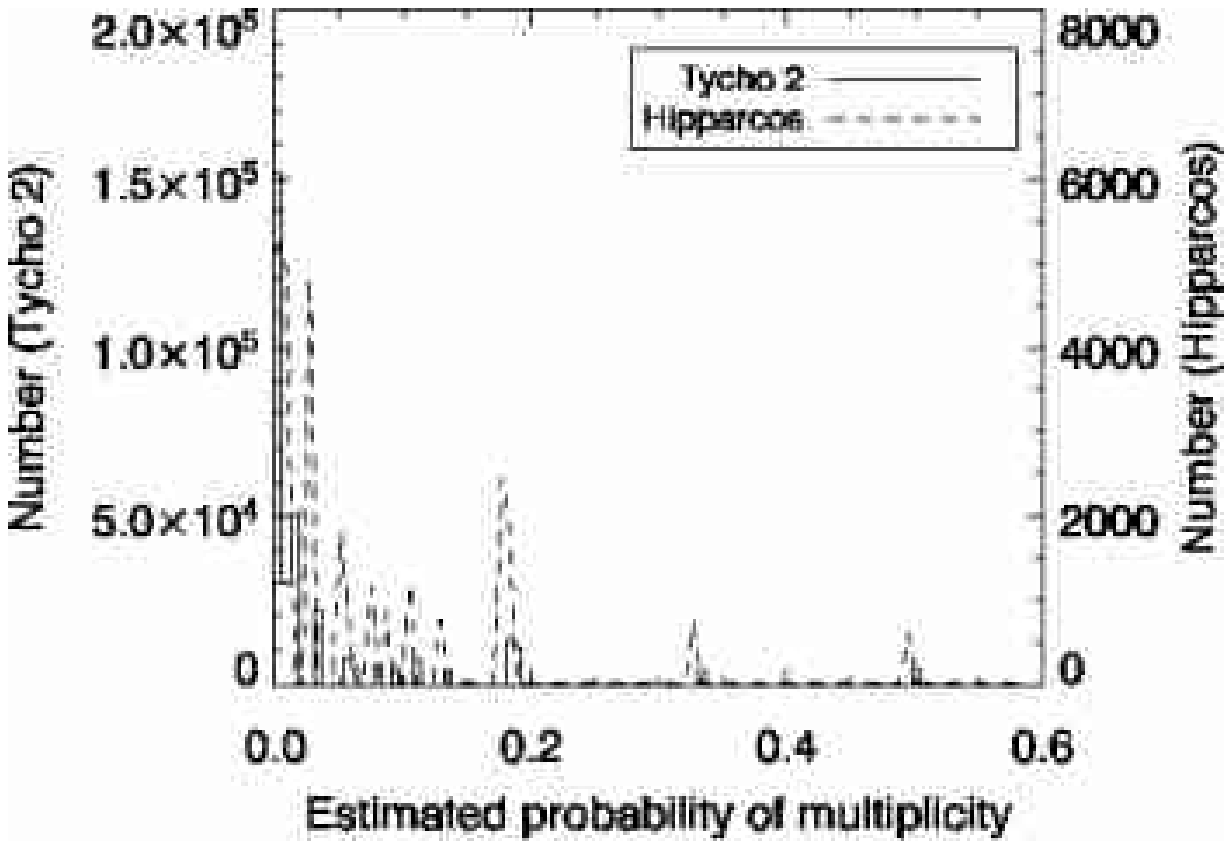}{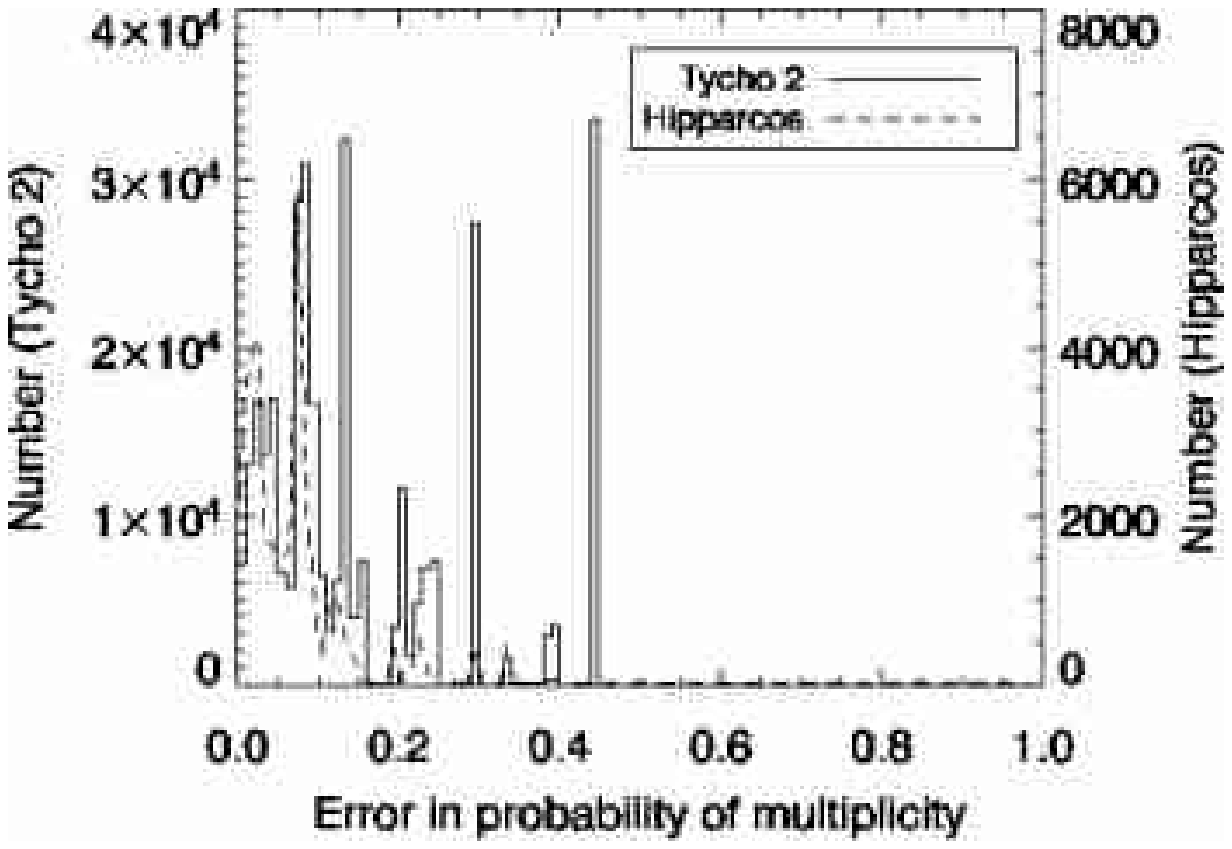}}
\centerline{\plottwo{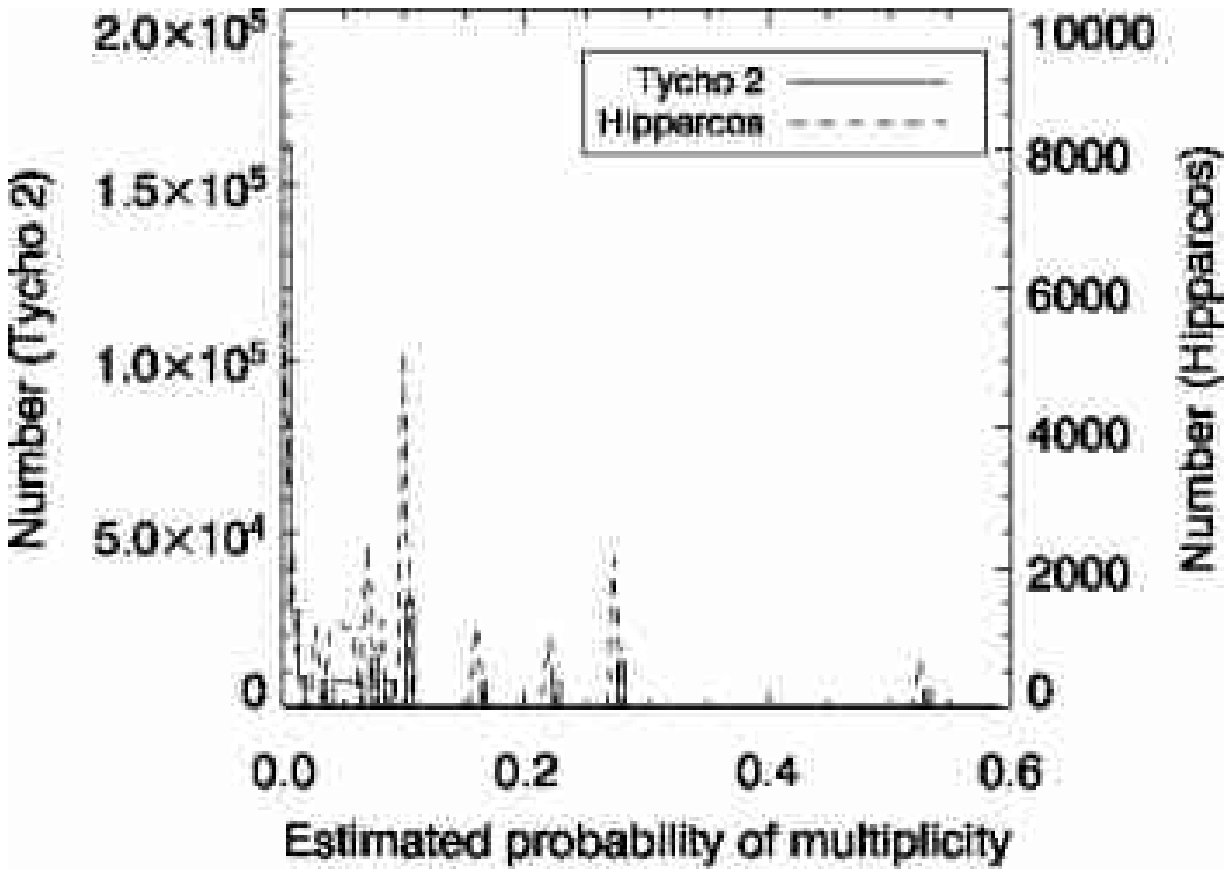}{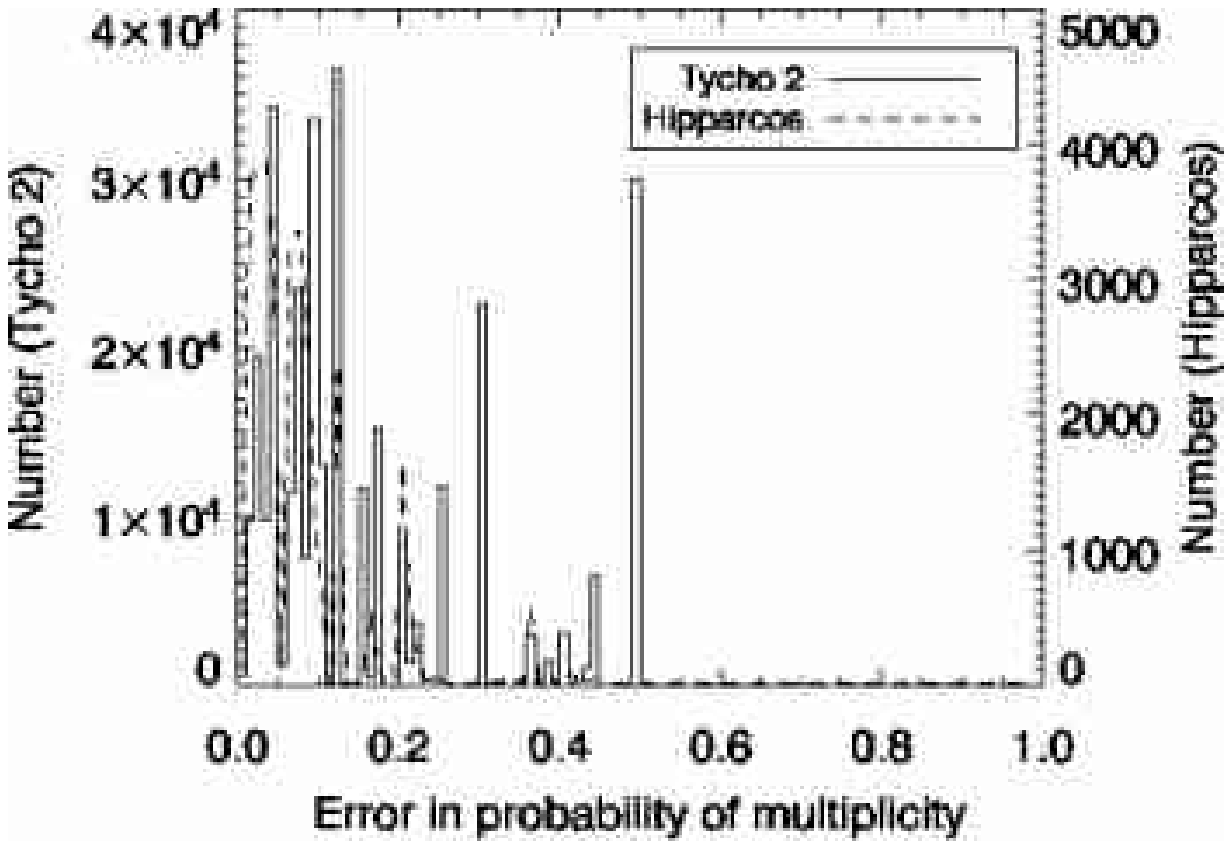}}
\caption{The solid lines are histograms for Tycho 2 stars in the dwarf pool; the dashed lines are histograms for the Hipparcos stars in the pool.  Error widths are calculated by
averaging the positive and negative intervals and are representative of a one-sigma error.  Top left:  Histogram of estimates of the probability of multiplicity for the dwarf pool
using a model that trained on a Kroupa, Tout, \& Gilmore (1993) IMF (see \S 3.4.1).  Top right:  Histogram of errors for this model.  Bottom left:  Histogram of estimates of the
probability of multiplicity for the dwarf pool using a model that trained on a modified Kroupa, Tout, \& Gilmore (1993) IMF (see \S 3.4.1).  This modified IMF has $\alpha = -2.35$
for $M > 1 \Msun\ $.  Bottom right:  Histogram of errors for this model.}\label{fig7}
\end{figure}

\begin{figure}[htb]
\centerline{\plottwo{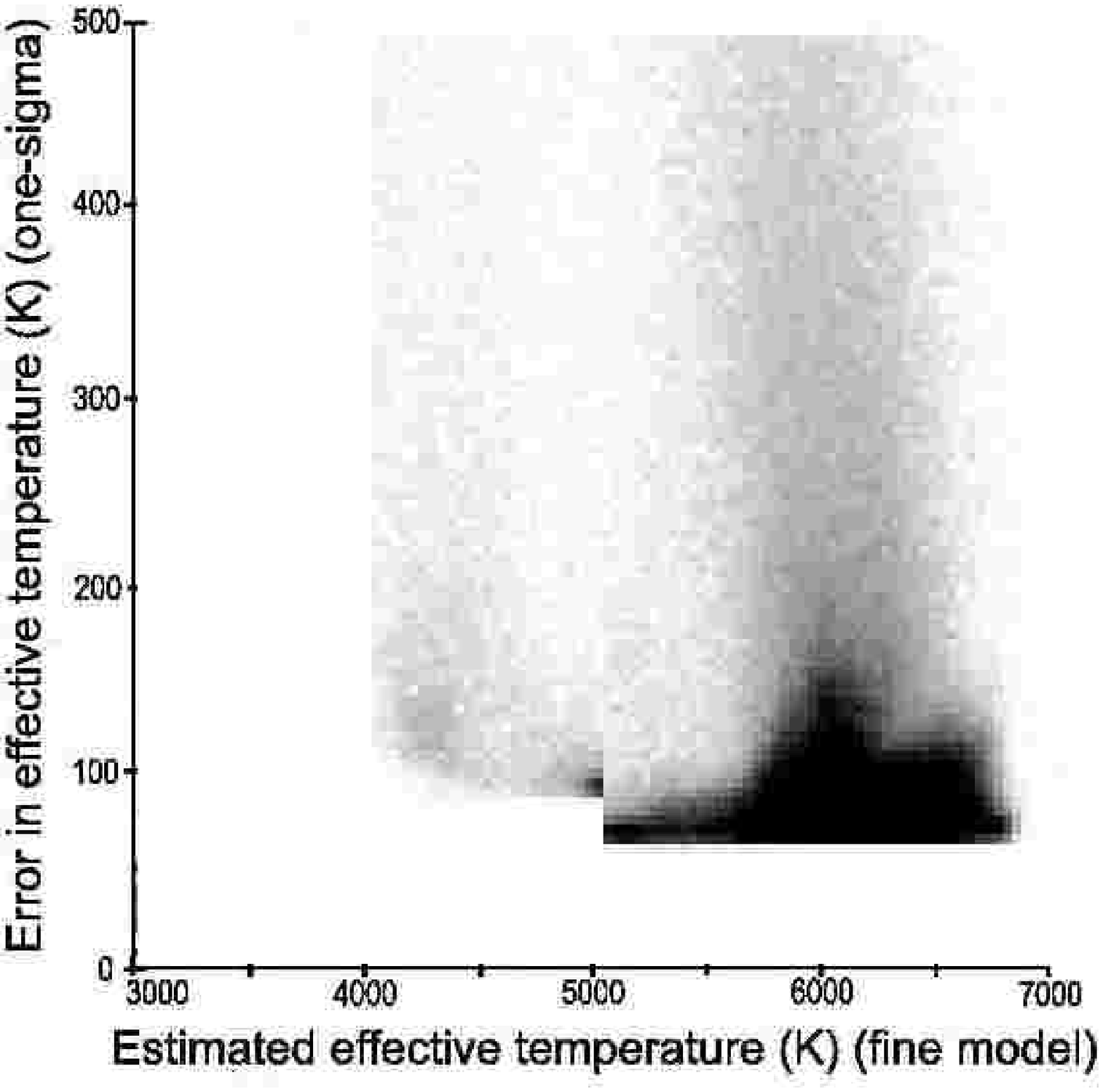}{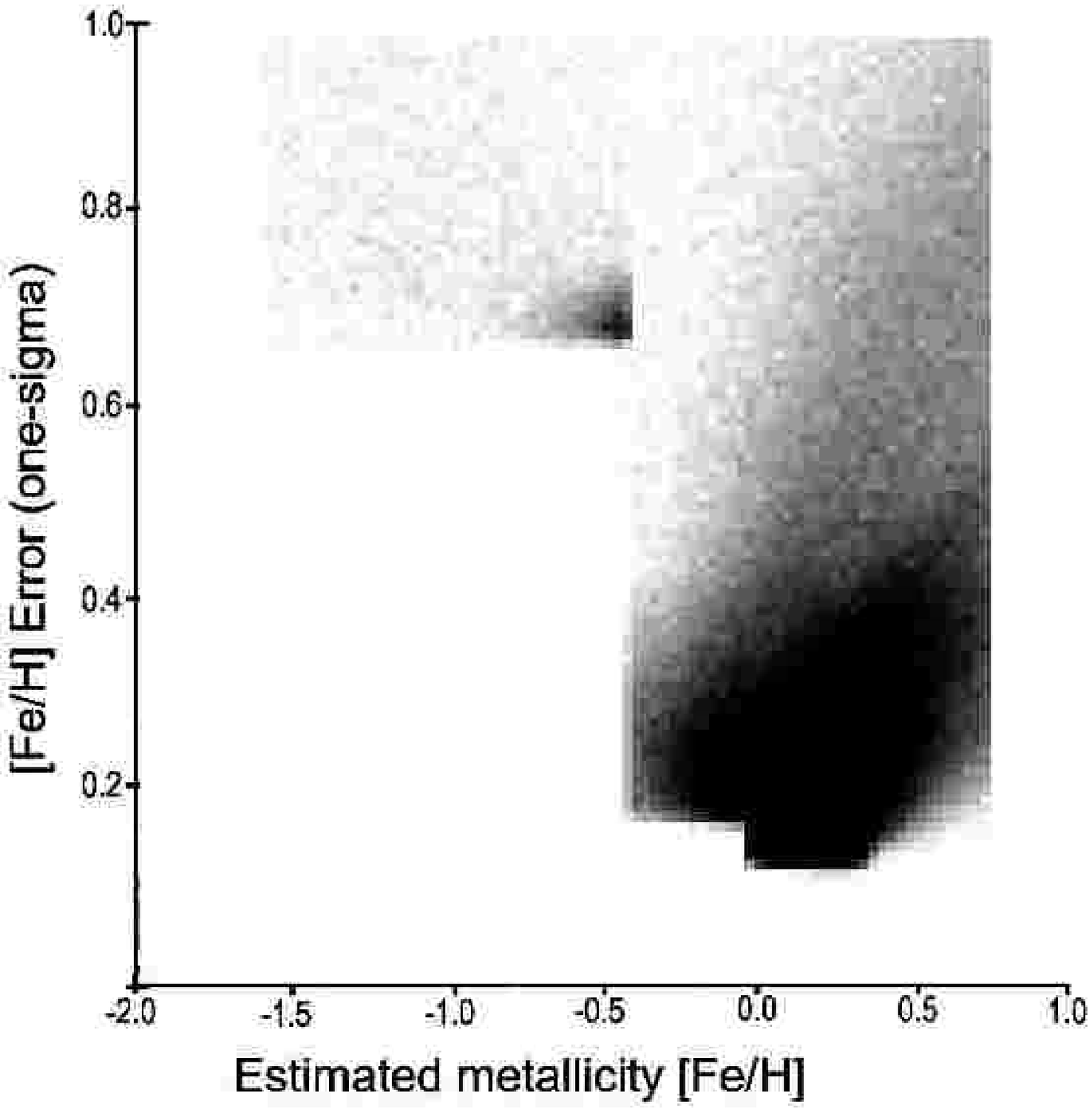}}
\centerline{\plottwo{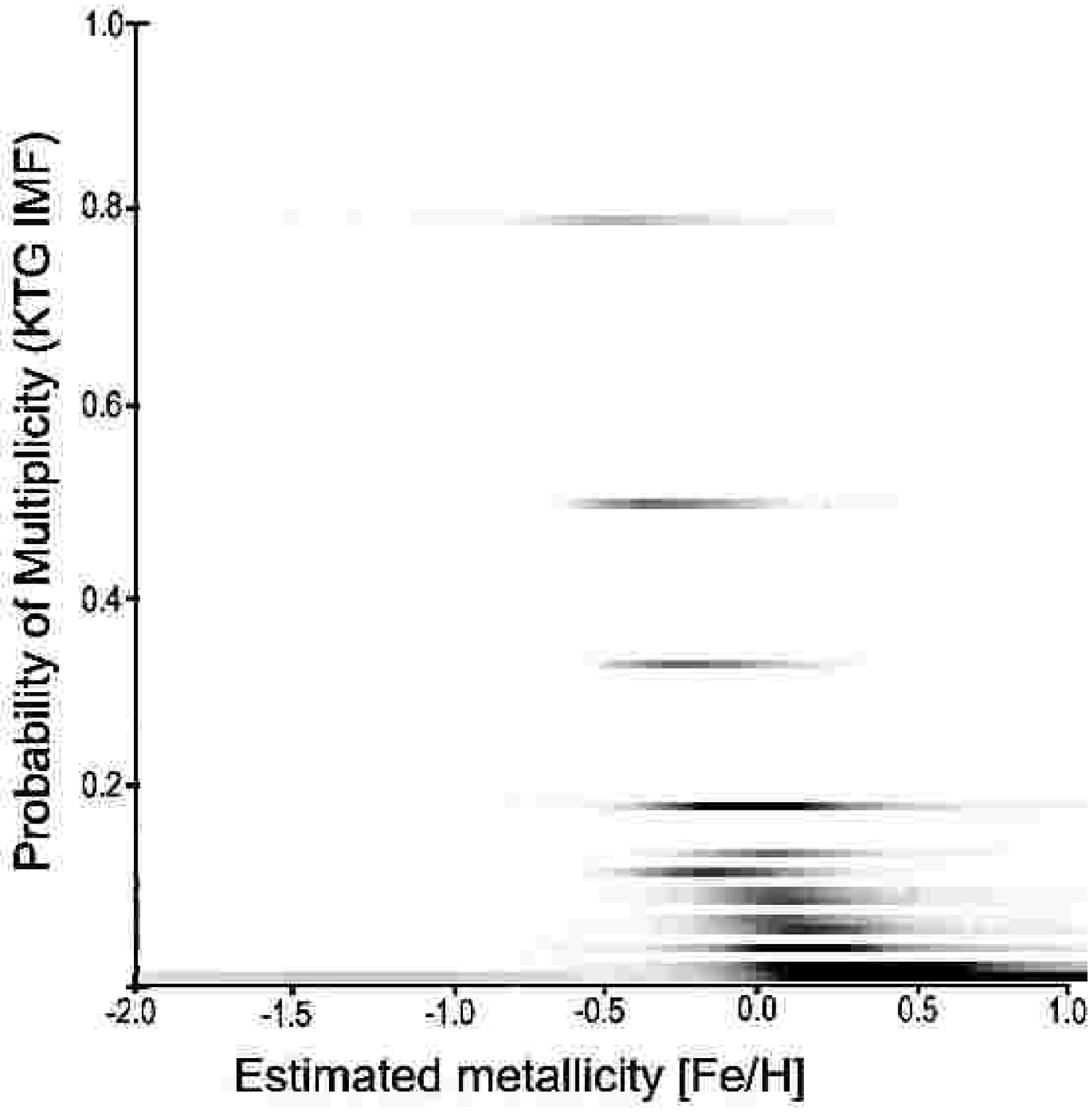}{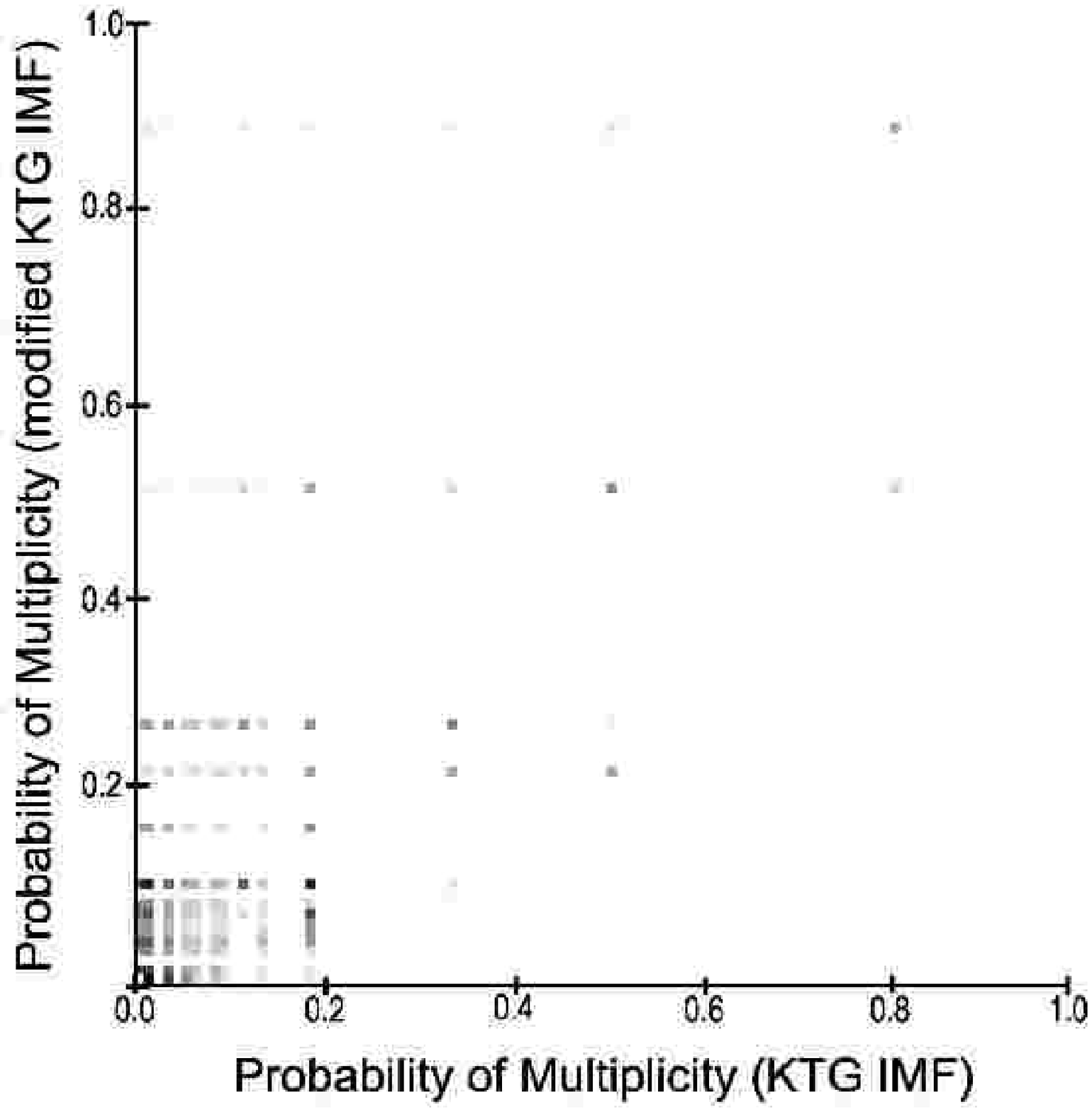}}
\caption{Density plots for Tycho dwarf pool.  Top left:  Plot of effective temperature from the ``fine'' temperature model versus temperature error.  Top right:  Plot of \FeH\
versus [Fe/H] error.  Bottom left:  Plot of \FeH\ versus estimate of probability of multiplicity.  Note the trend between multiplicity and metal-poor composition.  Bottom right: 
Direct comparison between the probability of multiplicity from two different models for the Tycho dwarf pool.  The horizontal axis is for the model trained on a Kroupa, Tout, and
Gilmore (1993) IMF and the vertical axis is for the model trained on a modified KTG IMF (see \S 3.4.2).}\label{fig8}
\end{figure}

We use the subset of Tycho 2 stars that fall in Hipparcos, a collection numbering 32,826 stars, to estimate giant/subgiant contamination.  Using the spectral types and luminosity classes available in Hipparcos, we find that contamination is low and that 88\% of the pool is composed of genuine F, G, and K type dwarfs.  2.6\% of the sample stars are giants and supergiants, 7.2\% are subgiants, and 2.0\% are other types of dwarfs.  Tycho is a magnitude limited sample that reaches fainter than Hipparcos, so we expect to find a greater dwarf/giant ratio in the rest of the dwarf pool; however, the contamination would likely be worse than that estimated from Hipparcos because of photometry error and the inadequacies of our simple reddening model.  

\section{ERROR}

Errors for all polynomial outputs have been estimated from histograms of residuals.  The errors quoted in this paper for each model define the one-sigma limits, or the regions within which $68\%$ of the training set errors fall.  The errors given in the published data set are functions of the model output values, calculated by binning the test set by model output and determining the $68\%$ interval for each bin.  The $68\%$ interval for each bin is calculated by sorting the residuals and counting inwards from the edge until $32\%$ of the set is reached.  These errors are found in the final step of the 3-way data split method (see \S 2).  The test set, which is the only untouched part of the entire training set, is used to generate unbiased residuals and errors.  Additional errors are added to these in quadrature to account for photometry errors, as described below.

Individual contributions to the published errors are as follows.  (1) The base error is intrinsic scatter due to physical processes.  For example, age variation is a source of this type of scatter in the \FeH\ models.  (2)  Modeling errors due to the insufficiency of simple polynomials to describe real physical trends are present.  (3)  Observational errors of the training parameter bias the results systematically.  (4) Misestimated errors for these parameters cause some stars to be mistakenly weighted, affecting the final errors.  (5) Systematic errors are caused by severe differences between the training set (Hipparcos) and the application set (Tycho), as described in \S 2.4.  (6) Contamination in the application set may occur by stars for which the models are not optimized.  For example, the \FeH\ and binarity models are only accurate for dwarfs; giant contaminants in the dwarf pool will have incorrect estimates of \FeH\ and multiplicity probability.  (7) Photometry errors appear to be the dominant source of stochastic error for dim stars.  (8) Lastly, temperature ``misbinning'' error occurs because of our choice of spline fits for some of the models.  Each of these models consists of several polynomials for different temperature bins.  For each star, the temperature is estimated first so that it can be assigned to the correct polynomial for \FeH\ and binarity.  There is a chance that the star is assigned to the wrong bin (i.e., mis-typing) if there is error in the temperature estimate.  

Errors due to (1) are generally dominated by the other sources of error, except for the brightest stars ($m_v < 9$).  Errors due to (2) are reduced by permitting higher order fits and using unconnected splines in our fitting routine.  We find improvements in (3) by using the Valenti \& Fischer (2005) SME set, a uniform collection of bright dwarfs for which accurate stellar atmospheric parameters have been measured with HIRES at Keck.  Compared to the Valenti \& Fischer set, the Cayrel de Strobel (1997) stellar atmospheric parameters are highly nonuniform.  Error (4) is reduced by assigning overall uncertainties that reflect this nonuniformity.  Error (5) may be reduced by only considering brighter stars ($m_v < 10.0$) in Tycho 2, as these stars more resemble Hipparcos and represent its particular biases and selection effects.  The contamination difficulties described in (6) are not represented in the published error bars, although our attempt at quantifying them (see \S 4) reveals low amounts of subgiant/giant/spectral type contamination in the dwarf pool.

We find that error due to item (7), photometry noise, is the chief source of error for the dwarf set.  For metallicity, we find that $92.8\%$ of stars brighter than $V_T = 10.0$ that satisfy $-2.0 < \FeH\ < 0.6$ have a one-sigma error of $\sim 0.13-0.3$ dex in [Fe/H].  Stars dimmer than $V_T = 10.0$ quickly become dominated by photometry errors.  $98.5\%$ of stars brighter than $V_T = 9.0$ that satisfy $-2.0 < \FeH\ < 0.6$ have a one-sigma error of $\sim 0.13-0.2$ dex in [Fe/H].  $28\%$ of the Tycho dwarf pool stars that satisfy $-2.0 < \FeH\ < 0.6$ and do not fall in Hipparcos have \FeH\ one-sigma error better than $0.3$ dex, or $\sim10^5$ stars.  

We address the misbinning error referred to in item (8) by manually adjusting the values output by the model.  The probability that a star is assigned to the wrong bin is known if Gaussian statistics are assumed and the temperature error is known.  A more accurate answer is obtained by evaluating the polynomials in the surrounding bins and combining them with the original result using a weighted sum.  The weights are given by the probabilities that the star falls in a particular bin.  The scatter errors are also combined in this manner (after the photometry error contribution is added to each).  A few Gaussian integrals give the general result
\begin{equation}
B_{best} = \frac{1}{\sqrt{2\pi}\sigma_T}\left(B_1 \int_{T_{est} - T_1}^{\infty}e^{-\frac{T^2}{2\sigma_T^2}}dT + B_2 \int_{T_1 - T_{est}}^{T_2 - T_{est}}e^{-\frac{T^2}{2\sigma_T^2}}dT + B_3 \int_{T_2 - T_{est}}^{\infty}e^{-\frac{T^2}{2\sigma_T^2}}dT\right),
\end{equation}
where $T_{est}$ is the estimated temperature of the star, $\sigma_T$ is the error in this value, $B_1, B_2, B_3$ are the estimates of the training parameter using the different polynomials ($B_1$ is for the cooler bin, $B_2$ is for the bin that $T_{est}$ lies within, and $B_3$ is for the hotter), and $T_1$ and $T_2$ are the boundaries between bins with $T_1 < T_{est} < T_2.$  The errors are combined with the same equation, substituting $\sigma_{B_i}$ for $B_i.$  This post-processing is a ``pseudo-connection'' for our unconnected spline models.  This processing is not performed on the training set stars during polynomial construction because the temperature errors are extremely small compared to the width of the temperature bins ($\sigma_T < 0.1 (T_2 - T_1)$).  

For all stellar parameters, the quoted error for each individual star includes the Gaussian photometry error as propagated through the polynomials.  We have added this propagated error in quadrature with the model error to produce complete error estimates.  The scatter errors given in the abstract are the best case errors, i.e., they do not include photometry or misbinning error estimates. 

\section{DISCUSSION}

\subsection{Improvements over Past Studies}

The stellar relationship between \FeH\ and UV flux is familiar, and indeed our tests with U photometry and \FeH\ have been highly successful.  U data is not widely available, however, so it is both fortunate and interesting that optical and IR colors together provide a good substitute.  The reasons for our success are manifold.  First, past models have relied on training sets like Cayrel de Strobel (1997), which contains \FeH\ estimates from hundreds of authors employing different methods and instruments.  We estimate that the internal consistency of these types of sets is on the order of 0.15 dex; using such compilations prevents model accuracies better than this threshold.  Our \FeH\ models, however, train on large amounts of uniform data that are taken on a single instrument and are reduced with a single pipeline (Valenti \& Fischer 2005).  In addition, the HIRES spectra used here are of sufficiently high resolution to remove rapid rotators and spectroscopic binaries.  Nearly every Tycho/2MASS flux used from this set is produced by a non-multiple star.  Subgiants have been isolated and removed from the training set.  

A further improvement is the use of IR data for the entire training set.  Although the \FeH\ models are more sensitive to $B_T$ and $V_T$ than IR magnitudes, $B-V$ color alone is degenerate with temperature.  In the $B-V$ CMD, increasing \FeH\ moves stars to the lower-right along the main sequence; thus, for instance, metal-rich G dwarfs are easy to confuse with metal-poor K dwarfs.  This difficulty has been encountered before with broadband \FeH\ polynomials (Flynn and Morell 1997).  $V-K$, on the other hand, is more sensitive to temperature than \FeH\ and effectively breaks the ambiguity.  Thus, combining multi-wavelength photometry is key to developing these polynomial fits, in agreement with several good fits of broadband IR fluxes to \FeH\ found in Kotoneva et al. (2002).  Finally, the use of the flexible fitting routine described in \S 2 quickens the process, permitting many flavors of fitting polynomials to be checked in rapid succession.  

We find that G stars possess colors with more abundance sensitivity than other dwarfs, in agreement with Lenz et al. (1998).  In this past study, the authors numerically propagated Kurucz (1991) synthetic spectra through the SDSS filters to summarize the possibilities of extracting abundance, intrinsic luminosity, and temperature information from intermediate-band photometry.  We have largely broken the ambiguity between luminosity and \FeH\ mentioned in Straizys (1985) by using spline functions rather than simple polynomials.  Straizys stresses the difficulty of using short wavelength photometry (e.g., $B_T$) at large distances due to reddening, which we tackle using reddening corrections for stars away from the Galactic plane.  Our \FeH\ models show good performance for metal-rich stars, complementing several models in the literature that use Stromgren narrowband photometry (Twarog 1980, Schuster and Nissan 1989a, Rocha-Pinto and Maciel 1996, Favata et al. 1997, Martell and Laughlin 2002, Twarog et al. 2002).  This improvement is due wholly to the good metal-rich sampling in the Valenti \& Fischer (2005) set.  We find that $\sigma_{[Fe/H]}$ is as small as $+0.114/-0.0807$ dex for bright metal-rich stars ($-0.067 <$ \FeH\ $ < 0.317$, $V < 9.0$).

\subsection{The Utility of Tycho for Radial Velocity Surveys}

We consider the suitability of a given star as being likely to harbor a Hot Jupiter type planet (Schneider 1996, Mayor et al. 1997).  For this purpose we suggest that a figure of merit be used to rank the Tycho 2 stars.  This figure of merit would be a function of the fundamental stellar properties calculated here, designed to isolate stars that are more likely to possess detectable Hot Jupiters according to known selection effects and biases.  Potential targets must have (1) surface temperatures between 4500 and 7000 K, (2) $d < 100$ pc, (3) no close binary companions, and (4) $\FeH\  > 0.2$ dex.  This last requirement relies on evidence that the presence of planets correlates with host metallicity (Fischer \& Valenti 2005).

Our broadband photometric estimates of \FeH\ have already been used to accurately filter metal-poor stars from radial-velocity target lists.  Low-resolution spectroscopy has shown that $60\%$ of bright FGK stars flagged as metal-rich (\FeH\ $> 0.2$) by the broadband models above truly satisfy this criterion (Robinson et al. 2005).  Additional stars not in the Valenti \& Fischer (2005) set that have been screened at Keck with HIRES have metallicities that agree with their broadband estimates within 0.1-0.15 dex (Fischer et al. 2005b)

We recommend that the Tycho 2 catalog stars be considered for radial velocity survey candidacy.  The $\it{uvby}$ data set of bright stars (Hauck and Mermilliod 1998) has traditionally been the reservoir of targets for radial velocity surveys, as \FeH\ polynomials of $\it{uvby}$
photometry may reach accuracies of $0.1$ dex (Martell and Laughlin 2002).  Alternatively, U broadband photometry has been used to estimate \FeH\ through UV excess (Carney 1979, Cameron 1985, Karaali et al. 2003).  Unfortunately, few currently untargeted stars have U, $\it{uvby},$ or other narrowband photometry available.  If \FeH\ estimates from optical and IR broadband photometry prove to be as robust as traditional U and $\it{uvby}$ estimates have been, mining existing catalogs like Tycho 2 is within reason.  Several difficulties in adopting this strategy include a significant reduction in brightness and a lack of distance estimates.  This latter deficiency prevents complete removal of subgiant/giant contaminants in the dwarf pools.  This may be addressed with low-resolution spectroscopy on small telescopes to serve as a filtering highway between the lowest level (broadband filtering) and the highest level (large high resolution telescopes).  The utility of this strategy is currently being proven for the N2K project (Fischer et al. 2005a and 2005b, Robinson et al. 2005).  

As for the overall reduction in brightness associated with mining Tycho 2, the arduousness of monitoring dimmer objects is not insurmountable, and future large-scale surveys will require this change of strategy.  The current trend of repeatedly observing the same set of stars in search of ever lower mass objects may not continue indefinitely; at some desired $v\,sin(i)$ accuracy the random line of sight components of gas velocity on a target star will overwhelm its mean orbit velocity and increase the measurement cost/benefit ratio beyond acceptable values.  Large-scale surveys like N2K (Fischer et al. 2005a), most notably, will help distinguish planet formation and migration scenarios, determine any trends with age or formation environment, and increase the likelihood of finding transiting planets.  

\subsection{Future Improvements}

Using photometric \Logg\ to perform dwarf/giant discrimination is important to the future of isolating dwarf and/or giant pools.  For nearby stars, reduced proper motion has been the classic method of discrimination, but deep surveys like the Sloan Digital Sky Survey have quickly outstripped the astrometric state of the art.  We find a degeneracy between \Logg\ and temperature for all dwarfs and giants, which is broken when we generate several polynomials with different temperature ranges (refer to the spline formulation described in \S 2).  Our experiments with \Logg\ models have shown that colors alone are sufficient to isolate pools of cool dwarfs $(T < 4000\; K)$ with less than $50\%$ contamination by red giants.  Previous tests (Dohm-Palmer et al. 2000, Helmi et al. 2003) suggest that this pass rate is reasonable.  Good dwarf/giant discrimination performance has been found in the Spaghetti survey (Morrison et al. 2001) and in other searches for halo giants (e.g., Majewski et al. 2000), which utilize modified versions of the Washington photometry (Canterna 1976, Geisler 1984) that have a strong surface gravity sensitivity.  Unfortunately, the number of stars with this photometry available is small compared to the number with SDSS fluxes.  

Overall, however, our experiments with broadband \Logg\ models have not been favorable.  The reasons for this are as follows:  (1) The physical processes that differentiate dwarfs from giants in photometry vary widely as a function of surface temperature.  A single polynomial or even a spline cannot be expected to capture all possible effects.  (2) Entire groups of stars are underrepresented in the Cayrel de Strobel (1997) / Valenti \& Fischer (2005) training set, namely blue giants and cool red dwarfs.  The expected number of cool red dwarfs in the Tycho 2 set is certainly a small percentage of the total number as well.  In addition, past studies have shown that \Logg\ varies only by small amounts in hot dwarfs and is weakly dependent on luminosity type (Newberg and Yanny 1998) and that cool red dwarfs are notoriously difficult to differentiate from K giants (Lenz et al. 1998).  There is some surface gravity information in photometry, however; for instance, deeper molecular lines in the IR bands of red dwarfs may be manifest in the photometry.  To improve the performance of a \Logg\ model, it will be necessary to increase the number of K giants and cool red dwarfs with good spectroscopic measurements in the training sets.

Apart from \Logg\, we expect to make improvements in the models that decrease the effects of photometry error.  As mentioned in \S 2.1, the models published here are optimized for stars with very good photometry.  This is sufficient for sorting target lists for N2K, which only operates on bright stars.  The applications enumerated in the section below will require photometry-optimized models, which use a $\chi^2$ statistic modified to include the effect of Gaussian photometry error.  

\subsection{Further Applications}

Applications for the Tycho 2 set include searching for \FeH\ gradients with Galactic radius (Nordstrom et al. 2004), searching for common proper motion groups with uniform abundances (e.g., Montes et al. 2001, L\'opez-Santiago et al. 2001), and sifting between star formation scenarios to best reproduce the distribution of these ``moving groups.''  Photometric abundance models may also be applied to extremely distant, possibly extragalactic stars that are too faint for targeted spectroscopy, permitting chemical evolution studies of our close satellites or even Andromeda (utilizing adaptive optics to get IR fluxes) in the low surface brightness regions.  

A few potential applications of our models for deeper data sets include the correlation of abundance gradients with galactic location, the search for particular populations in the halo and evidence for past mergers events, differentiating thick and thin disk populations with broadband \FeH\ alone, and sifting among Galactic star formation scenarios using this information.  The key to using these models on distant objects is developing an accurate binarity proxy that searches for uncharacteristic IR brightening and removes binaries with intermediate mass ratios.  Our own binarity model is theoretically capable of isolating large pools of stars in which binary contamination is low.  It is not necessary to remove binaries that do not have intermediate mass ratios (i.e., \mrat\ $ < 1.25$ or \mrat\ $ > 3$) because (1) systems with unity mass ratio consist of stars with similar abundances and colors, which would not mislead broadband temperature or metallicity models and (2) systems with stars of vastly different absolute magnitude are dominated in color by the primary.  

Galactic structure analyses utilizing position-dependent star counts (see, e.g., Bahcall \& Soneira 1981 and 1984, Reid \& Majewski 1993, Infante 1986, Infante 1994, Chen et al. 1999, Chen et al. 2001) can be built upon immediately with the current data set and fitting framework.  We have produced a Sloan/2MASS overlap list of 800,000 stars, for which we have generated Johnson/Cousins $B,V$ magnitudes with the Smith et al. (2002) conversion polynomials.  Choosing either G dwarfs or K giants as a tracer population, we have applied our abundance and binarity models to the overlap set and have obtained star counts in several \FeH\ bins.  We have used photometric \FeH\ alone to distinguish the galactic thick disk from the thin disk with this data.  The conversion from the SDSS $\it{u'g'r'i'z'}$ system (Fukugita et al. 1996) to broadband effectively reduces the number of resolution elements available, so we intend to ultimately transfer the Valenti \& Fischer (2005) set to the SDSS filter system through photometric telescope observations.  Barring contamination and reddening uncertainties, such a conversion will enable unprecedented galactic structure and chemical evolution studies to be performed out to large disk heights.  

\subsection{Summary}

We have used an extensive training set (Valenti \& Fischer 2005) of excellent spectroscopic measurements of atmospheric parameters to produce models of fundamental stellar parameters.  A least-squares and/or absolute deviation minimization procedure assisted us in finding spline fits between properties like [Fe/H], $T_{\rm eff}$, distance, and binarity and $B_T$, $V_T$, J, H, Ks fluxes and proper motion.  We have used the well-documented three-way data split statistical method to choose best-fit model parameters and estimate unbiased errors.  All data products are publicly available at the Astrophysical Journal Supplement (website to be specified upon publication).  The \FeH\ model achieves remarkable accuracy for metal-rich stars and will be crucial for sorting target lists for future large-scale radial velocity planet searches like N2K (Fischer et al. 2005a).  The binarity model, which to our knowledge is the first of its kind in the literature, will be useful for sorting target lists as well.  A total of 100,000 FGK dwarfs in the published dwarf pool are bright stars that retain $0.13-0.3$ dex \FeH\ accuracy and $80-100$ K temperature accuracy, but are absent from Hipparcos.

\section{Acknowledgements}

The authors wish to thank Tim Castellano for assistance with Hipparcos, Greg Spear for observing assistance, K.L. Tah for data mining experience, Chris McCarthy for sharing giant/dwarf discrimination methods, and Connie Rockosi for literature search support.  S.M.A., S.E.R, and J.S. acknowledge support by the National Science Foundation through
Graduate Research Fellowships.  Additional support for this research was provided by the NASA Origins of Solar Systems Program through grant NNG04GN30G to G.L.

{}

\end{document}